\documentclass[12pt,preprint]{aastex}

\shorttitle{Nearby debris disk systems with high fractional luminosity}
\shortauthors{Mo\'or et al.}

\begin{document}


\title{Nearby debris disk systems with high fractional luminosity reconsidered}

\author{A. Mo\'or\altaffilmark{1}, P. \'Abrah\'am\altaffilmark{1}, 
A. Derekas\altaffilmark{2,3}, Cs. Kiss\altaffilmark{1}, L.L. Kiss\altaffilmark{2}, 
D. Apai\altaffilmark{4,5}, C. Grady\altaffilmark{6,7}, Th.~Henning\altaffilmark{8}}
\affil{1) Konkoly Observatory of the Hungarian Academy of Sciences, PO Box 67, H-1525 Budapest, Hungary}
\affil{2) School of Physics A28, University of Sydney, NSW 2006, Australia}
\affil{3) School of Physics, Department of Astrophysics and Optics, University of New South Wales, NSW 2052, Australia}
\affil{4) Steward Observatory, The University of Arizona, 933 N. Cherry Avenue, Tucson, AZ 85721, USA}
\affil{5) NASA Astrobiology Institute}
\affil{6) Eureka Scientific, 2452 Delmer Street Suite 100, Oakland, CA 94602-3017}
\affil{7) Exo-Planets and Stellar Astrophysics Laboratory, Exploration of the Universe Division, 
NASA Goddard Space Flight Center, Code 667, Greenbelt, MD 20771}
\affil{8) Max-Planck-Institut f\"ur Astronomie, K\"onigstuhl 17, 69117 Heidelberg, Germany}
\begin{abstract}
By searching the IRAS and ISO databases we compiled a 
list of 60 debris 
disks which exhibit the highest fractional luminosity values ($f_d>10^{-4}$)
in the vicinity of the Sun ($d<120$\,pc). Eleven out of these 60 systems are new discoveries. 
Special care was taken to exclude bogus disks from the sample. 
We computed the fractional luminosity values using 
available IRAS, ISO, and Spitzer data, and analysed
the galactic space velocities of the objects. The results revealed that stars with disks 
of high fractional luminosity often belong to 
young stellar kinematic groups, providing an opportunity to obtain improved age 
estimates for these systems. We found that practically all disks with 
$f_d>5{\times}10^{-4}$ are younger than 100\,Myr. 
The distribution of the disks in the fractional luminosity versus age 
diagram indicates that (1) the number of old systems with high $f_d$ is
 lower than was claimed before; (2) there exist many relatively 
young disks of moderate fractional luminosity; and (3) comparing the observations with
a current theoretical model of debris disk evolution a general good agreement could be found.

\end{abstract}





\section{Introduction} \label{intro}

One of the major discoveries of the IRAS mission was that main-sequence stars 
may exhibit excess emission at infrared (IR) wavelengths ("Vega-phenomenon",
Aumann et al. 1984). Systematic searches in the IRAS catalogues  
(Backman \& Paresce 1993; Mannings \& Barlow 1998; Silverstone 2000; 
Zuckerman \& Song 2004a,
and references therein)  revealed that $\sim 15$\% of  main-sequence stars show
infrared excess \citep{plets99}. It was suggested already after the first discovery 
that the excess can be attributed to thermal emission of dust
confined into a circumstellar disk \citep{aumann84}. The existence of such
\it debris disks \rm was first confirmed by the coronographic observation of scattered light
from the $\beta$ Pic system \citep{smith84}.  Subsequent imaging of specific
systems at mid-infrared  (e.g. HR 4796A, Koerner et al. 1998) and submillimeter
wavelengths  (e.g. $\epsilon$ Eri, Holland et al. 1998) supported this picture.

The possibility that debris disks might evolve over time was first mentioned
by \citet{backman87} who proposed that  disks around Vega, $\epsilon$~Eridani, and
Fomalhaut could be more evolved analogs of the $\beta$ Pic system. Using
submillimeter measurements of the best known Vega-like stars \citet{holland98}
showed  that the derived dust mass of the disks decay with stellar age
as a power-law. Subsequent studies with the \it Infrared Space
Observatory \rm (ISO, Kessler et al. 1996)  demonstrated that debris disks are
more common around young stars ($t < 400$\,Myr) than around old ones and there
is a trend for older debris disks to be less massive than younger ones
\citep{habing01,sprangler01,sil00}.  The evolutionary picture was
further refined by \citet{decin03}, who reinvestigated the ISO results and
revised the stellar age estimates. Recently \citet{rieke05} presented a
survey of A-type stars, performed at 24$\rm \mu m$ with the Multiband Imaging
Photometer for Spitzer (MIPS, Rieke et al. 2004) onboard the {\it Spitzer Space
Telescope} \citep{werner04}, and observed a general decay on a timescale of
150\,Myr.

A new generation of theoretical models has been developed to explain
the temporal evolution of debris disks \citep{kenyon02,dominik03,kenyon04}.
These models take into account the fact that the destruction timescales of
dust  grains orbiting main-sequence stars are significantly shorter than the 
age of  the central star, therefore the observed dust grains in a debris disk
must  be continuously replenished \citep{backman93}. Collisional erosion of
minor bodies in exosolar analogs of our Solar System or the sublimation of
comets are the best explanations for the  replenishment process
\citep{harper84,backman93}. Most recent models link the temporal
evolution of debris disks to the formation and erosion of planetesimals \citep{kenyon02,kenyon04}.

The amount of dust in a debris disk is usually characterised by its fractional
luminosity, $f_d$, defined as the ratio of integrated infrared excess of the
disk to the bolometric luminosity of the star.  Despite the general
evolutionary trend described above, fractional luminosity values of individual
systems were found to show a large spread of $10^{-6}\lesssim f_d \lesssim 10^{-3}$ at almost any age \citep{decin03}.
\citet{rieke05} also found large variations of the IR excess around A-type stars within each age group.
They emphasized the role of individual collisional events between large planetesimals as one of the possible explanations
of their result. 

Particularly interesting is the relatively high number of older systems (t$\gtrsim$500\,Myr) with
high fractional luminosity values ($f_d \simeq 10^{-3}$). These systems pose a
serious challenge even to the new generation of theoretical models.  A
qualitative explanation for the existence of debris disks with large 
$f_d$ values around old stars was suggested by \citet{dominik03}. According to
their theory, different planetesimal disks become  active debris
systems at different ages, because of the delayed onset of collisional
cascades.

The presence of relatively old systems with high fractional luminosity,
however, might partly be an observational artifact, since several factors
could bias the distribution of debris disks on the $f_d$ vs. $age$
diagram. Some stars have been nominated as Vega-candidates due to erroneous
infrared photometric measurements, confusion by background sources, or the
presence of extended nebulosity  (where the IR emission is of interstellar
rather than of circumstellar origin). In order to obtain a reliable picture of
debris disk evolution these misidentifications ("bogus disks") have to be found and discarded.
The actual positions of confirmed debris disks in the $f_d$
vs. $age$ diagram might also be biased by measurement errors in the far-infrared
photometry and even more significantly uncertainties in the age determination.

In this paper we study a sample of debris disks 
which exhibit the highest fractional luminosity values
in the solar neighbourhood. Setting a threshold value of 
$f_d = 10^{-4}$ and a distance limit of 120\,pc we compiled a
list of 60 disks, and  performed an accurate
determination of their infrared excess
using IRAS, ISO, and Spitzer data (Sect.\,2).
In Sect.\,3 we present improved age estimates for a large
number of stars, derived e.g., by determining their membership in young moving groups.
In Sect.\,4 we analyze the distribution of the disks in the $f_d$ vs. $age$ diagram and 
find a significant fraction of our sample to be younger than was previously thought.
%We test the hypothesis that a significant fraction of our sample of debris 
%disks are much younger than was previously thought;  
%and analyze the distribution of the disks in
%the $f_d$ vs. $age$ diagram (Sect.\,4). 
Our conclusions are summarised in
Sect.\,5. In Appendix\,A. we list bogus disks identified in our work. 
Appendix\,B. lists new members of young moving groups
discovered in the present study.

\section{Sample selection}

We created the input list for this study by: (1) identifying debris disk candidates
in the IRAS and ISO databases;
(2) rejecting bogus debris disks and suspicious objects; 
(3) computing infrared fractional luminosity values and selecting disks with $f_d > 10^{-4}$.

%The input list for our study was created in three steps: (1) we collected
%debris disk candidates by searching the {\it IRAS} and {\it ISO} databases; 
%(2) rejected bogus debris disks and suspicious objects from the sample; (3)
%computed infrared fractional luminosity values and select disks with $f_d >
%10^{-4}$.  

\subsection{Searching the IRAS catalogues for stars with infrared excess} 
\label{selection}

With the aim of compiling a list of main-sequence stars with IR excess, we
made a systematic search in  the IRAS Faint Source Survey Catalog (FSC, Moshir et
al. 1989) and in the IRAS Serendipitous Survey Catalog 
(SSC, Kleinmann et al. 1986).  In order to reduce source confusion,
our survey was confined to $|b| \geq 10^\circ$ Galactic latitudes.  We selected
all infrared sources with at least moderate flux quality at 25 or 60$\rm \mu m$,
and their positions were correlated with entries from the Hipparcos Catalogue
(ESA 1997) and the Tycho-2 Spectral Type Catalogue \citep{wright03}. 
Positional coincidences within 30\,arcsec were extracted.  In order to assure
that the selected objects are not giant stars, the luminosity class was
constrained to IV-V in the Hipparcos, and to V in the Tycho catalogue.
We also included several objects whose luminosity class was not available in the
Hipparcos catalogue but their absolute magnitudes indicated a main-sequence
evolutionary phase. Since the infrared excess from early B-type stars might be
due to free-free emission \citep{zuckerman01} the sample was limited to
spectral types later than B9 in both catalogues.  Our query is similar to that
of \citet{sil00} but -- since we also considered the IRAS 25$\rm \mu m$ band --
it is also sensitive to stars with excess from warmer dust disk.

Following the principles of the method by \citet{plets99}, for each selected star  we predicted 
the far-infrared flux density of the stellar photosphere 
using the $\rm K_s$-band magnitude (or V-band, when good quality $\rm K_s$ photometry was not available) and 
the B--V color index. $\rm K_s$-band photometry was drawn from the Two Micron All Sky Survey (2MASS) catalog \citep{cutri03}, 
V magnitudes and B--V color indices were taken from the Hipparcos and Tycho Catalogues. 
As a first step a photospheric 25$\rm \mu m$ flux density was derived from
the $\rm K_s$ magnitude and the B--V color of the star using the collection of stellar
model predictions by M.\,Cohen and P.\,Hammersley (available on the
ISO Data Centre home page).
Then, color relationships predicting the photospheric flux ratios between 
25$\rm \mu m$ 
and a selection of IRAS, ISO, and Spitzer photometric bands were also derived from the same stellar models.
The average accuracy of the predicted far-infrared fluxes is estimated to be
around 4\% when computed from the $K_s$-magnitudes, and 
8\% when computed from V-magnitudes. 

In order to compute IR excess values the predicted photospheric flux densities were subtracted from the 
measured flux densities in each IRAS band.
In principle the IRAS fluxes have to be color corrected since the shape of the spectral energy distribution 
of the system usually differs from the 
$F_{\nu}{\sim}{\nu}^{-1}$ reference spectrum 
(this spectral shape was assumed while the flux densities quoted in the 
IRAS catalogues were derived from the detector in-band powers).
Since the true spectrum of the system is not known a priori,  
we decided to multiply the predicted photospheric fluxes --  
rather than dividing the IRAS flux densities -- with 
color correction factors appropriate for a stellar photosphere
(IRAS Explanatory Supplement, Beichman et al. 1988). 
The significance level of the infrared excess was calculated in each 
photometric band with the following formula:
\begin{equation}
S_{excess} = \frac{F_{meas} - F_{pred}}{\sqrt{{{\delta}{F^2_{meas}}} + {{\delta}{F^2_{pred}}}}} 
\end{equation}   
where ${\delta}{F_{meas}}$ is the quoted uncertainty in the FSC or SSC,
and  ${\delta}{F_{pred}}$ is the uncertainty of the prediction
described above. When $S_{excess}$ was greater than 3 either in 
the 25 or 60$\rm \mu m$ bands, the object was selected as a 
excess candidate star. Applying the above criteria we identified in total 355 excess candidate stars in the IRAS 
databases.

\subsection{ISO-based selection of stars with infrared excesses} \label{ISOselection}

In a second step 
the IRAS-based list was supplemented with excess stars selected from
the ISO databases.
The Vega-phenomenon was a key programme for ISO (see Sect.~\ref{intro}) and a number of stars have been observed with ISOPHOT, the
onboard photometer \citep{lemke96}. We collected all ISOPHOT observations of normal stars from different observers 
performed in mini-map, sparse-map or staring
mode (a detailed description of these observing modes is given in The ISO Handbook Vol.~IV, Laureijs et al. 2003), 
and performed a homogeneous re-evalution of the whole sample (\'Abrah\'am et al. 2003; \'Abrah\'am et al., 2006, in prep.).
For details of the data analysis and criteria for candidate excess stars, see \'Abrah\'am et al. (2006, in prep.).
We note that most selected candidates have already been published by the original
observers \citep{decin00,habing01,sprangler01,sil00}, but due to their different 
processing schemes the published flux densities cannot be directly merged for a homogeneous
catalogue. 
The merged IRAS- and ISO-based lists includes altogether 364 IR excess stars. 

\subsection{Rejection of suspicious objects} \label{rejection}

Since our goal is to compile a list of debris disks, we excluded all  known
young stellar objects (e.g. T Tauri or Herbig Ae/Be stars) which harbour
protoplanetary disks. The sample could also be contaminated by source
confusion: due to the low spatial resolution of IRAS at far-IR
wavelengths, many of the positional coincidences between a star and a far-IR
source could be bogus and the far-IR emission is related to a foreground or
background object.  

For part of the sample (110 stars)  higher spatial resolution infrared
maps are available, obtained either by the ISOPHOT or the MIPS instrument. We downloaded 
ISOPHOT data from the ISO Data Archive (IDA) and
processed with the Phot Interactive Analysis (PIA) version 10.0
\citep{gabriel97}. 
MIPS Basic Calibrated Data (BCD) files were downloaded from the Spitzer Science Center (SSC) 
data archive.
These latter products are composed of 2-dimensional FITS image files which 
included all general calibrations and corrections for MIPS detectors \citep{gordon05}.
In each case these data were coadded and corrected for array distortions with the
SSC MOPEX (MOsaicking and Point source Extraction, Makovoz \& Marleau 2005)
software. Bad data flagged in the BCD mask files, as well as permanently damaged pixels flagged in the 
static pixel mask file were ignored during the data combination.
Output mosaics had pixels with size of 2.5$''$ at 24$\mu m$ and 4$''$ at 70$\mu m$.
SSC MOPEX/APEX software package was used to detect sources and determine their positions on the final maps.

The positions of the infrared sources were determined on the ISO and Spitzer maps
and objects whose coordinates differed from the optical position 
(and in some cases coincided with a nearby background object), and/or
associated with extended nebulosity, were discarded from the list.
For positional discrepancy the threshold value was set to half of the 
width of a point source's footprint. In the case of ISOPHOT the large pixel size dominated 
the footprint and in the 60--100$\mu m$ range the threshold was 23$''$.
In the case of MIPS arrays footprint was defined by the telescope Point Spread Function and we adopted 
threshold values of 3$''$ and 9$''$ at 24 and 70$\mu m$, respectively. The absolute pointing uncertainty 
was less than these values for both satellites.
In total 24 disk candidates were dropped from the list.

When neither ISO nor Spitzer maps were available we 
made an attempt to filter out bogus disks by assuming that an object is possibly affected by source confusion if:
\begin{itemize}
\item a known galaxy or evolved star (OH/IR source, Mira variable) is located within 1 arcmin of the IRAS position;   
\item a source included in the IRAS Small Scale Structure Catalog \citep{helou} or in the 2MASS Extended Source Catalog
\citep{jarrett00}
is located within 1 arcmin of the IRAS position; 
\item a 2MASS source with an excess in the $\rm K_s$ band (identified in the $\rm H-K_s$ vs. $\rm J-H$ diagram in comparison with the 
locus of the main-sequence and taken into account the reddening path) 
is located within 1 arcmin of the IRAS position; 
\item the 60-to-100$\rm \mu m$ flux ratio of the candidate source 
resembles the color of infrared cirrus 
$(\frac{F_{60}}{F_{100}} < 0.25$, which correspond blackbody temperatures lower than $33$\,K ).
At least moderate flux quality flags were required in both IRAS bands.
\end{itemize}
These cases were also discarded from our list of debris disk candidates (48 objects).

In an earlier study, \citet{kalas02} used 
high-angular
resolution coronographic observations at optical wavelengths 
and found cases where the far-infrared excess observed by IRAS
was of interstellar -- rather than circumstellar -- origin ("Pleiades-phenomenon"), leading to false 
entries in the Vega-candidate lists. They also suggested that a significant
fraction of Vega-candidates beyond the Local Bubble might be bogus, since the star illuminates nearby interstellar matter
rather than a circumstellar disk.
After checking the positions of our sources projected on recent maps of the Local Bubble \citep{lall03},
we discarded all objects situated in the wall of the bubble or beyond.
The wall was defined as the isocontours corresponding 
to the 50\,$\rm m\AA$ NaI D2-line equivalent widths in the maps.
In practice nearly all of our sources beyond 120\,pc were removed, while within this radius only a few were dropped.
Thus we defined a maximum distance limit of 120\,pc for our stars, constructing a 
nearly complete volume-limited sample.

\subsection{The list of disks with high fractional luminosity} \label{finallist}

In order to compute fractional luminosity values for each candidate star, 
we constructed spectral energy distributions 
by combining infrared fluxes from the 
FSC, SSC, ISOPHOT (re-evaluated by us, see \'Abrah\'am et al., 2006, in prep.) and 
additional MIPS and submillimeter fluxes from the literature. 
The excess above the predicted photosphere was fitted by a single temperature modified blackbody, 
where the emissivity was assumed to vary as $1-\exp{[-(\lambda_0/\lambda)^\beta]}$ where $\lambda_0$
was set to 100$\mu$m (see e.g. Williams et al., 2004).
 We fixed ${\beta}$ equal to 1, which is a typical 
value in the case of debris systems \citep{dent00}.
If the excess was detected at one wavelength only, 
we adopted a modified blackbody whose peak (in $F_{\nu}$) coincided 
with that single wavelength. 
From the fitted spectral shape color correction factors were computed and applied to the data.
Then again a modified blackbody was fitted resulting in new color correction factors, and this procedure was repeated
until the color correction factors converged.
Finally, the fractional dust luminosity was calculated as $f_d = L_{IR} / L_{bol}$. 
In order to estimate the uncertainties on our fractional luminosity values we performed a Monte Carlo simulation.
We added Gaussian noise to the photometric data points using their quoted 1$\sigma$ photometric errors and then 
recomputed the fractional dust luminosities. Formal uncertainties of the predicted theoretical photospheric
fluxes were also taken into account.
Final uncertainties were derived as the standard deviation of these values after 1000 repetitions. 
We note that these values include only random uncertainties; systematic errors due to e.g. limited wavelength coverage 
are not taken into account.

\citet{artym96} argued that debris disks are confined to $f_d < 10^{-2}$ and sources with higher fractional luminosity 
probably contain a significant amount of gas (e.g. T Tau and Herbig Ae/Be stars, "transition" objects). 
Therefore we excluded objects with $f_d > 10^{-2}$ from our sample. 
Then the remaining sample was sorted by decreasing $f_d$ values and stars with $f_d > 10^{-4}$ (60 stars) were taken for the 
further analysis
presented in this paper. 
%The threshold value of $10^{-4}$ is somewhat arbitrary,
%nevertheless it defines a list whose size is a compromise between having a
%statistically large sample and being able to analyse the observations of each
%star individually.

Basic stellar parameters for these 60 objects, as well as derived fractional luminosities and
their uncertainties, are presented in Table~\ref{tab1}.
Infrared data used in our analysis, including both the original flux values as listed in the catalogues or 
provided by our reduction algorithm, as well as corrected
fluxes, where color correction was applied, are given in Table~\ref{tabflux}. The table also contains photospheric
flux predictions for the specific wavelengths. 
Inspecting the flux density values obtained by different instruments at the same wavelength (e.g. IRAS and ISOPHOT 
at 60$\mu$m) one 
finds discrepancies which may arise e.g. from the different beams and different calibration strategies of the 
instruments. Comparing the IRAS and ISOPHOT flux values in Table~\ref{tabflux} a 
general good agreement within 1$\sigma$ was found, with no deviations above 
the 3$\sigma$ limit (the uncertainty $\sigma$ was computed as the quadratic sum of quoted uncertainties 
from the two instruments). 
We also compared our ISOPHOT flux densities with the results of earlier evaluations of the same observations in the literature.
In most cases the results were consistent (except HD\,10647 and HD\,53143 where Decin et al. (2000) derived significantly higher 
values at 60$\mu$m; the probable explanation is that we used a more advanced version, V10.0, of the PIA software).

As was discussed in Sect.~\ref{selection}, during the analysis we rejected several systems 
as bogus disks or suspicious objects. Those rejected stars which were previously proposed to harbour debris disks 
in the literature and would have been included in our final list (on the basis of their quoted fractional luminosity in the original
paper) are presented in Appendix\,A. together with a brief 
description of the reason of rejection.

\section{Age determination} \label{kinage}

\subsection{Membership in young moving groups} \label{mg}

Age determination for main-sequence field stars is challenging, and sometimes
results in very uncertain values. Ages of open cluster members, however, can be
estimated more accurately e.g. by fitting their main-sequence locus in the
color-magnitude diagram  with theoretical isochrones or by determining the location of the "lithium depletion edge" 
in the cluster and comparing it with the predictions of theoretical evolutionary models. A number of young
clusters ($\alpha$ Per, Pleiades, Hyades etc.) have been dated so far 
\citep{meynet93,stauffer98}. Similarly, the ages of young
stellar kinematic groups, discovered mainly in recent years, are relatively
well determined (e.g. Zuckerman \& Song 2004b). It was a very important result 
that several stars with the
strongest infrared excess turned out to be members of such moving groups, 
and in some cases the ages of these stars had to be
revised significantly (e.g. the case of $\beta$ Pic, Barrado
y Navascu\'es et al. 1999). In order to obtain more reliable ages for our sample, 
we performed a systematic investigation of  the possible
relationship between our excess stars and nearby kinematic moving groups,
stellar associations or open clusters.     

A common method to decide whether an object belongs to a moving group is to compare its Galactic space velocity components
with the mean velocity components of the group. In order to compute the space velocity for the stars in Table~\ref{tab1}, 
we collected 
parallaxes and proper motions from the Hipparcos and Tycho-2 catalogues. When accurate parallax information was not 
available, a photometric distance was adopted.
  
  Radial velocities were taken from the literature (see Table~\ref{tab3} for references), or from our own observations. 
The new observations were carried out with the 2.3m ANU-telescope at the Siding
Spring Observatory, Australia, on 11 nights between 21 March and 22 August 2005.
The spectra were taken with the Double Beam Spectrograph using a 1200mm$^{-1}$
grating in the red arm. The recorded spectra covered 1000 \AA\ between 
5800--6800 \AA, with a dispersion of 0.55 \AA px$^{-1}$. This leads to a nominal
resolution of about 1 \AA. The exposure time ranged between 30\,s and 200\,s
depending on the brightness of the target and the weather conditions. We
obtained on average 4--7 spectra for each star. Since all the target stars are
bright objects (V$<$10 mag), we could easily reach S/N$\sim$150--200 for every
spectrum. All spectra were reduced with standard tasks in IRAF\footnote{IRAF is
distributed by the National Optical Astronomy Observatories, which are operated
by the Association of Universities for Research in Astronomy, Inc., under
cooperative agreement with the National Science Foundation.}. Reduction
consisted of bias and flat field corrections, aperture extraction, wavelength
calibration and continuum normalization. We did not attempt flux calibration
because the conditions were often non-photometric and the main aim was to measure
radial velocities. Radial velocities were determined by cross-correlation, using the IRAF task
{\it fxcor}, choosing HD~187691 as a stable IAU velocity standard. The
cross-correlated region was 100 \AA\ centered on the H$\alpha$ line, which is
far the strongest spectral feature in our range. The finally adopted velocities
were calculated as simple mean values of the individual  measurements. Our
experiences have shown that the typical measurement errors were about 4--7
km~s$^{-1}$ per point, so that the mean values have $\pm$1--3 km~s$^{-1}$
standard deviations. These were adopted as the uncertainties shown in Table~\ref{tab3}.

In the calculation of the Galactic space velocity we used a right-handed coordinate system (U is positive towards 
the Galactic centre, V is positive in the direction of galactic rotation and W is positive towards the North galactic pole) and 
followed the general recipe described in
"The Hipparcos and Tycho Catalogues"
(ESA 1997). The computed Galactic space velocity components and their uncertainties are given in Table~\ref{tab3}.

Table~\ref{tab2} summarizes the basic properties of the relevant moving groups and associations within 120\,pc from 
the Sun. 
The probability that star $i$ is a member of moving group $j$, can be computed by:

\begin{equation}
P_{ij} = \exp \left( -\left[\frac{(U_i - U_j)^2} {2 ({\sigma_{U_i}}^2 + {\sigma_{U_j}}^2)} + 
                \frac{(V_i - V_j)^2} {2 ({\sigma_{V_i}}^2 + {\sigma_{V_j}}^2)} +
		\frac{(W_i - W_j)^2} {2 ({\sigma_{W_i}}^2 + {\sigma_{W_j}}^2)} \right] \right)
\end{equation}
where $U_i,V_i,W_i$ and $\sigma_{U_i},\sigma_{V_i},\sigma_{W_i}$ are the  
Galactic space velocity components of the star and their uncertainties,
respectively, while $U_j,V_j,W_j$ and $\sigma_{U_j},\sigma_{V_j},\sigma_{W_j }$
are the mean Galactic space velocity components of the specific kinematic group
as well as the corresponding errors. In this formula we assumed that the
velocity distribution within a group is Gaussian. When
$\sigma_{U_j},\sigma_{V_j},\sigma_{W_j }$ parameters were not available in the
literature, we computed them from the velocity dispersion of known
members around the mean. Our newly calculated mean values were always
consistent  with those from the literature within the uncertainties. In those
few cases where no sufficient membership information could be found  in the
literature, we adopted $\sigma_{U_j} = \sigma_{V_j} =\sigma_{W_j } = \rm
2$\,km~s$^{-1}$ (a characteristic value in the previous cases). 

Probability values for each star with respect to each group were computed.
Then we checked the resulting $P_{ij}$ values for objects already assigned 
to a group in the literature. The numbers 
spread in the range of $0.2 < P < 1.0$, therefore we set $P = 0.2$ as a lower limit
for the new moving group member candidates as well. Stars assigned to any group above this threshold 
were further checked by comparing their 3-dimensional space location with the
volume occupied by the group 
(most groups are rather confined in space). 
There were a few stars which could be assigned to both the Tucana-Horologium and GAYA2 associations; these cases are 
analyzed in Appendix\,B. Table~\ref{tab3} presents the final assignments between stars and kinematic groups. 

From our sample of 60 objects, 26 sources could be linked to stellar associations; 
13 of them are new members identified in the present study. For 10 stars out of these 13, age estimates are available in
the literature. For these ten objects we directly compare our age estimates with 
previous values in Table~\ref{tabage}. In most cases ages derived in our study
are younger than the earlier values. The discrepancy is particularly obvious in the case of ages derived from isochrone
fitting (for example Nordstr\"om et al. 2004). However, this problem is not related only to the
present study. For example, HD\,105 has a moving group age of $30^{+10}_{-20}$\,Myr \citep{mamajek04} while 
\citet{nord04} quoted $8600^{+4000}_{-3800}$\,Myr from isochrone fitting. Similar objects in our sample 
are HD\,25457 
($50-100$\,Myr from moving group, Zuckerman \& Song 2004b and $4000^{+1200}_{-2100}$\,Myr from isochrones, 
Nordstr\"om et al. 2004); HD\,164249 
($12^{+8}_{-4}$\,Myr, Song et al. 2003 and $2200^{+1200}_{-1800}$\,Myr, Nordstr\"om et al. 2004); HD\,181327 
( $12^{+8}_{-4}$\,Myr, Song et al. 2003 and $1300^{+1000}_{-1300}$\,Myr, Nordstr\"om et al. 2004). 
The age uncertainty related to isochrone fitting might arise from the lack of information on whether the star is 
in the pre-main-sequence phase of its evolution or is an evolved object above the main-sequence. 
In ambiguous cases we always adopted age estimates derived from stellar kinematic group membership.

\subsection{Statistical age estimates for the disk sample}

For stars not assigned to any moving groups other age estimation methods are needed. 
Before focusing on individual systems, in this subsection we analyze what 
can be learned about the 
age distribution of our sample of debris disk systems.  

\subsubsection{Distribution of the excess stars in the velocity space}

The distribution of the derived Galactic space velocities (Table~\ref{tab3})
are displayed in Fig.\,1a-b. Overplotted is the box occupied by young disk
population stars defined by \citep{leggett92} on the basis of a systematic study of Eggen (1989). 
The plots show that most stars from our
sample  belong to this population. This fact suggest that the majority of our
sample of stars from Table~\ref{tab1} are relatively young.

\subsubsection{Location of A-type stars on the CMD} \label{cmd}

For a sample of bright A-type stars \citet{jura98} demonstrated that three
objects with strong infrared excess (HR\,4796A, HD\,9672, $\rm \beta\,Pic$)
are  located close to the lower boundary of the distribution in the 
Color-Magnitude Diagram (CMD). Comparing their locations with the loci of the
youngest, nearby open clusters ($\alpha$\,Per,  IC2391, Pleiades),
\citet{lowrance00} argued that stars with strong infrared excess are typically
younger than these clusters. Adopting this idea we selected stars with B$-$V
ranging between -0.1 and 0.33  (corresponding mainly to A-type stars) from
Table~\ref{tab1}, and plotted them in the CMD of Fig.~\ref{figcmd}. In
addition we overplotted a volume-limited sample of A-type stars ($d<100$\,pc)
extracted from the Hipparcos Catalogue (it was requested that the parallax error
was less than 10\% and the B$-$V uncertainty was lower than 0.01\,mag).  For
comparison, members of the open clusters $\alpha$\,Per (80\,Myr) and Hyades
(600\,Myr) are marked.  The older cluster, Hyades, covers the upper part of the
distribution, while $\alpha$\,Per stars  are situated at lower absolute
magnitudes for the same color. Pleiades (100\,Myr, 
not plotted in the figure
for clarity) occupies the same region as $\alpha$\,Per (see Fig.\,3 in Lowrance
et al. 2000).

Figure~\ref{figcmd} shows that the majority of the objects from our sample of high
$f_d$ stars appear to be close to the lower boundary of the area occupied by A-stars,
and are located below the region of
$\alpha$\,Per and Pleiades, with some overlap. This suggests that the early
spectral type stars from our sample of high $f_d$ disks are young,
probably close to the zero age main-sequence (ZAMS), and they are presumably not older than 
 100\,Myr, the age of the Pleiades.

\subsection{Ages of individual objects} \label{agerest}

For those objects which could be assigned to one of the moving groups or
associations %in Table~\ref{tab2}% 
the age of that group as well as its
uncertainty was adopted (26 objects).
For a number of stars not associated with groups or with associations, 
age estimates could be found
in the literature (19 stars, for references see Table~\ref{tab1}). 
Sometimes literature data for a specific star scatter significantly; in these cases we 
adopted an age range which covers all quoted values and their uncertainties.   
When the literature search did not yield any dating, we made age estimates by plotting the
stars on the HR diagram and comparing their positions to isochrones. 
For stars with spectral types in the range B9-G5 the isochrone age was estimated 
following the general outline described by \citet{lachaume99}, using the Padova theoretical isochrones \citep{girardi00}. 
This method was applied to 13 objects. 

This isochrone method gives only upper limits for some A-type stars. 
%This is not surprising, since they are located close to the ZAMS in Fig.~\ref{figcmd}. 
As a best estimate for these stars (five cases), we 
adopted an upper limit of 100$\rm\,Myr$, consistent
with our results shown in Fig.~\ref{figcmd} and discussed in Section ~\ref{cmd}.
For stars of later spectral type there are some widely used age indicators, like the strength of the Ca\,II\,H\&K 
lines or the X-ray luminosity of the star.
In the case of HD\,121812 our age estimate is based on the former method, taking the measured value from \citet{strassmeier00} 
and using the calibration of \citet{lachaume99}.
HD\,130693 has a ROSAT counterpart and its X-ray luminosity of $\rm \log{L_x} = 29.7$\,erg~s$^{-1}$ was compared with the
X-ray luminosity distribution function of late-type members in different associations (see 
figure\,2 in Stelzer \& Neuh\"auser 2000), yielding an age range of $10-100$\,Myr.      
In Table~\ref{tab1} we summarize the age estimates for each object.

\section{Discussion} \label{discussion}

\subsection{Connection between debris disks and young moving groups} \label{connection}

From our sample of 60 main-sequence stars exhibiting strong infrared excess, 26 can be assigned to young stellar
kinematic groups. In order to test whether the frequency of stars belonging to young stellar kinematic groups is
similar in a general sample we determined the corresponding ratio within a volume-limited sample of normal stars.
%In order to place this result in context  
First we created this
sample by selecting stars from the Hipparcos catalogue using the following criteria: (1) 
they are closer as 120\,pc (the same volume-limit than in our sample);
(2) their Survey flag in the catalogue\footnote{Field H68 in the Hipparcos catalogue. 
'S' indicates that the entry is contained within the `survey',
which was the basic list of bright stars added to and merged with the total list of proposed stars, to 
provide a stellar sample almost complete to well-defined limits. 
The limiting magnitude was a function of the stars's spectral types and galactic latitude.} was set to "S"; 
(3) they have radial velocity measurement with uncertainty less than $\rm 5$\,km~s$^{-1}$.
The second condition guarantees that stars observed in various individual projects do not introduce
a bias in the analysis of the velocity distribution \citep{skuljan99}. 
In addition, 
\citet{binney97} noted that radial velocities are preferentially observed for high-proper-motion stars
which may cause a kinematical bias in our sample. Following the proposal of \citet{skuljan99} we constrained 
the sample for stars exhibiting low transverse velocity,
and excluded all stars with  $\rm v_t \geq 80$\,km~s$^{-1}$ in order to avoid this bias.
 
\begin{figure} 
\epsscale{.8}
\plotone{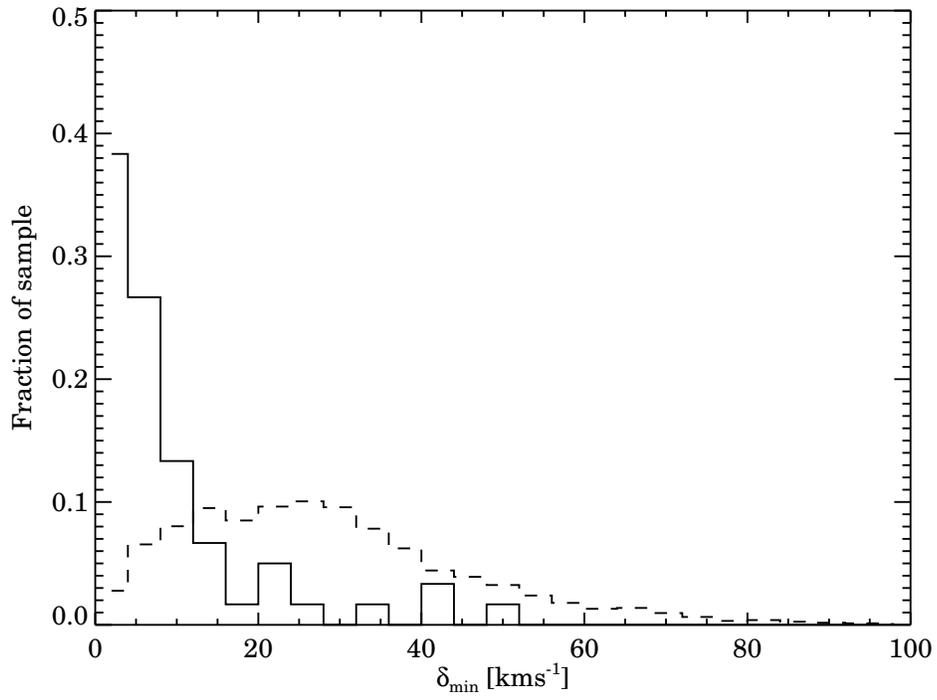}
\caption{Histogram of distances in the 3D velocity space to the closest moving group.
{\it Solid line:} stars from Table~\ref{tab3}; {\it dashed line:} the volume-limited Hipparcos sample (see Sect.\,4.1). \label{fighisto}}
\end{figure}

The query resulted in 7519 objects, for which we computed 
the UVW Galactic
space motion components. For each star in the two samples (the 60 debris systems in Table~\ref{tab1}, 
and the newly defined volume-limited stellar sample)
we determined the Euclidean distance in the 3D velocity space from the closest 
moving group, $\delta_{min}$. In this analysis we considered only groups younger than 150\,Myr. 
In Fig.~\ref{fighisto} we plotted the histograms of $\delta_{min}$ for the two samples. 
A two-sided Kolmogorov-Smirnov test shows that the two distributions
are different with a probability higher than 99.99\%.
%A two-sided Kolmogorov-Smirnov test shows the probability for the two distributions being part of the same sample is around 10 7.
This result indicates that debris systems of high infrared fractional luminosity 
are much more intimately linked to the nearby young stellar kinematic groups than the majority of normal
stars. 

\subsection{The relationship between fractional luminosity and age} \label{zsprop}

\begin{figure} 
\epsscale{.8}
\plotone{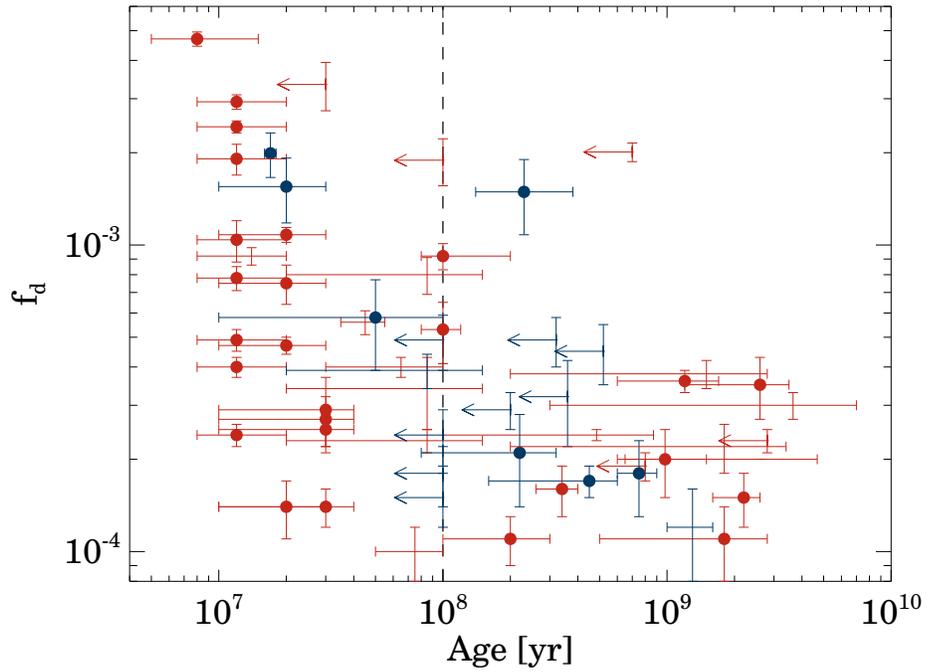}
\caption{Fractional luminosity of the infrared excess as a function of age.
Dashed line marks the threshold of 100\,Myr. Upper age limits are denoted by arrows
whose hats correspond to the uncertainty in $f_d$. When only an age range is known, no 
filled circle was plotted. Red symbols mark those debris systems whose existence was explicitely
confirmed by an instrument independent of IRAS (Sect.~\ref{zsprop}). \label{fdage1}}
\end{figure}

\citet{zs04b} hypothesised that stars with $f_d > 10^{-3}$ are younger than
100\,Myr, and therefore a high $f_d$ value can be used as an age indicator. This
proposal is in contradiction with the conclusion of \citet{decin03}, who
claimed the existence of high $f_d$  disks  around older stars. In order to
test which proposal is supported by our data, in Fig.~\ref{fdage1} we plotted
the distribution of ages as a function of the fractional luminosity $f_d$ from Table~\ref{tab1}.
 We plotted in red the debris disks whose presence was explicitely
confirmed by an instrument independent of IRAS (see Col.\,8 in Table~\ref{tab1}). 
The confirmation could be based e.g. on  high spatial resolution infrared images
(ISO/ISOPHOT, Spitzer/MIPS), or on mid-infrared spectra (Spitzer/IRS), or on
coronographic images (HST/ACS, HST/NICMOS). We found 43 
confirmed debris disks in total.  

Most data points with $f_d > 5 \times 10^{-4}$ fall below the age threshold of 100\,Myr (marked by a dashed line), 
while objects with lower $f_d$ show a larger spread in age. 
This trend can be recognized in the whole sample, but is especially clear in the confirmed subsample (red symbols). 
There is only one noteworthy case: HD\,121812 with an age of $230^{+150}_{-90}$\,Myr
exhibits fractional luminosity exceeding the $5 \times 10^{-4}$ threshold value. However, the 
presence of a debris disk around this star has not been confirmed independently of IRAS.  
On the basis of this result we conclude that 
-- according to the suggestion of
\citet{zs04b} -- the majority of debris disks with  $f_d > 5 \times 10^{-4}$
are younger than 100\,Myr, and high fractional luminosities can be used as an
indicator of youth. Nevertheless the opposite is not true, i.e. a low $f_d$ value is
not correlated with age, and in particular is not an indicator of antiquity.

  There is a growing list of debris disk systems which have been discovered by the sensitive 
detectors of the Spitzer Space Telescope \citep{beichman05a,bryden05,chen05a,chen05b,kim05,low05,meyer04,stauffer05,uzpen05}, 
and one may wonder whether these new 
observations support our previous conclusion. As a preliminary check we collected from
the cited papers all debris disks with $f_d > 5 \times 10^{-4}$. 
We used fractional luminosity values and age estimates as quoted in the papers. 
We found that all of these new disks discovered so far belong to 
the $\sim$16\,Myr old Lower Centaurus Crux subgroup of the Scorpius-Centaurus association (HD\,106906, HD\,113556, HD\,113766, HD\,114082, 
HD\,115600, HD\,117214; Chen et al. 2005a), or to the 
TWA (TWA\,7, TWA\,13A, TWA\,13B; Low et al. 2005), or to the star forming region RCW\,49 
(18 possible warm debris disks, Uzpen et al. 2005), which unambigously shows that these objects 
are young, in agreement with the conclusion of the present paper.

\subsection{Debris disk evolution and the cases of old systems} \label{dominik}

There are a number of models in the literature (see Sect.\ref{intro}) to describe the
temporal evolution of debris disks. In the following we compare 
our results with predictions.

\citet{dominik03} proposed a simple collisional model which assumes that all
dust grains in the debris disk  are produced in collisions between
planetesimals within a ring whose radius is constant during the whole
evolution. In collisional equilibrium -- when the dust production and
destruction rates are in balance -- the grain loss mechanism governs the amount
of dust visible in the system. If dust destruction is dominated by collisions
the fractional luminosity $f_d$ decreases proportionally to $t^{-1}$. If the
dust removal process is dominated by the Poynting-Robertson drag,  
$f_d \propto t^{-2}$. It is predicted that in
disks with $f_d > 10^{-4}$ the evolution is dominated by collisions
\citep{dominik03,wyatt05}.

\begin{figure} 
\epsscale{.8}
\plotone{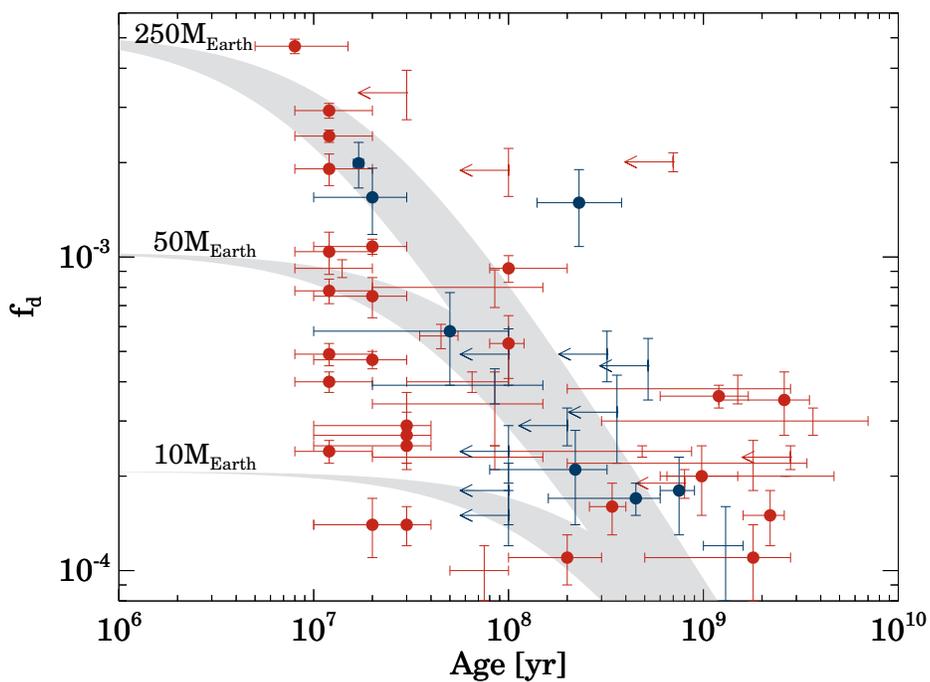}
\caption{Fractional luminosity of the infrared excess as a function of age. The shadowed bands 
mark the evolutionary models of \citet{dominik03} (for detailed model parameters see Sect.\ref{dominik}).
Upper age limits are denoted by arrows
whose hats correspond to the uncertainty in $f_d$. When only an age range is known, no 
filled circle was plotted.   \label{agefd}}
\end{figure}

Three families of disk evolution models, computed from eqs. 7, 35-40 of
\citet{dominik03}, are plotted as shadowed bands in Figure~\ref{agefd}.
The data points and symbols in this figure are identical to those in Fig.~\ref{fdage1}.
 The main differences between
the model families are related to disk mass: (a) $M_d = 10$\,$\rm M_{\oplus}$; 
(b) $M_d =$ 50\,$\rm M_{\oplus}$; and (c) $M_d =$250\,$\rm M_{\oplus}$.
The width of each band corresponds to a range in stellar mass 
from 0.5$\rm M_{\odot}$ to 3$\rm M_{\odot}$. Additional parameters are: 
the characteristic radius of the ring of planetesimals $r_{c} = 43\rm\,AU$; 
the radius of planetesimals $a_c = 10$\,km; the density of the planetesimal material 
$\rho_{c} = 1.5$\,g\,cm$^{-3}$ (proposed for icy bodies with small rocky component, Greenberg 1998; Kenyon 2002);
the size of the smallest visible grains $a_{vis} = 10\rm\mu m$ 
(taken from Jura et al. 2004); the absorption efficiency of the dust particles 
$Q_{abs} = 1$; and $\epsilon_0 = 226$ (defined in eq. 22 in Dominik \& Decin 2003).     

Figure~\ref{agefd} shows that the location of most stars on the evolutionary diagram can be 
explained by the models. 
The number of older stars exhibiting high $f_d$ values incompatible with 
the models is relatively low. Most of these disks are located in the 
$t \geq 10^9$\,yr and $f_d \lesssim 5\times10^{-4}$ area.
 A possible explanation for the origin of these stars was
proposed by \citet{dominik03} who assumed that different planetesimal disks 
become  active debris system at different ages because of the
delayed onset of the collisional cascade. Large collisional events may also increase the brightness of 
a debris disk temporarily \citep{rieke05}. 
Extraordinary events during the evolution like e.g. a proposed 'supercomet' in the HD\,69830 system (Beichman et al. 2005b)
cannot be excluded, too.
Nevertheless, the low number of systems incompatible with the models, especially at $f_d \gtrsim 5\times10^{-4}$, 
indicates that the above scenarios do not represent the main evolutionary trend.

It is important to note that a large spread in fractional luminosities ($10^{-4} < f_d < 5\times10^{-3}$)  can
be observed in the figure among young debris systems ($t < 100$\,Myr). 
This result resembles the
findings of \citet{rieke05} among A-type stars but somewhat contradicts to 
\citet{decin03} and \citet{dominik03} who found only few young stars with moderate or small
infrared excesses, and proposed that it might be related to the effect of stirring. 
A possible explanation of the large spread among young stars could be that the
initial conditions of the disks (especially initial disk mass) are far from being
homogeneous.

\section{Conclusions} \label{conclusion}

We searched the IRAS and ISO databases and compiled a list 
of debris 
disks exhibiting the highest fractional luminosity values ($f_d>10^{-4}$)
in the vicinity of the Sun ($d<120$\,pc). 
Utilizing high-resolution far-infrared maps we attempted to 
exclude bogus disks from the sample. 
The fractional luminosity value for each disk was recomputed using
available IRAS, ISO, and Spitzer data, and analysed
the galactic space velocities of the objects
as well as the distribution of the disks on the fractional luminosity 
versus age diagram. Our results are summarized as follows:
\begin{enumerate}
\item We compiled a list of 60 debris disk systems of high fractional luminosity. 
Eleven of them are new discoveries and 4 out of these 11 have been confirmed by Spitzer 
observations;
\item Disks with high fractional luminosity often belong to 
young stellar kinematic groups, providing an opportunity to obtain improved age 
estimates for these disks;
\item Practically all objects with $f_d>5{\times}10^{-4}$  
are younger than 100\,Myr;
\item  The number of old systems with high $f_d$ seems to be lower than was claimed before, mainly as a consequence
of the age revision in connection to the young stellar kinematic groups;
\item There exist many young disks of moderate fractional luminosity;
\item Comparing the theoretical evolutionary model of \citet{dominik03} with the observations
in the $f_d$ vs. $age$ diagram good general agreement was found. 

%The extension of the model
%to explain the presence of old stars with high fractional luminosity disks was not needed.
\end{enumerate}

\section{Acknowledgments}

We are grateful to the anonymous referee for his/her comments which improved the paper.

This research has made use of the IRAS and Hipparcos Catalogs 
(ESA, 1997),
as well as the SIMBAD database and the VizieR tool operated by 
CDS, Strasbourg, France.
Two Micron All Sky Survey (2MASS) is a joint project of the University of 
Massachusetts and the Infrared Processing and Analysis Center/California 
Institute of Technology, funded by the National Aeronautics and Space 
Administration and the National Science Foundation.

This work is based on observations made with the Spitzer Space  
Telescope, which is operated by the Jet Propulsion Laboratory,  
California Institute of Technology under a contract with NASA.  
Support for this work was provided by NASA through an award issued by  
JPL/Caltech.

This material is partly based upon work supported by the National  
Aeronautics and
Space Administration through the NASA Astrobiology Institute under
Cooperative Agreement No. CAN-02-OSS-02 issued through the
Office of Space Science.
%This work is based in part on observations made with the Spitzer Space 
%Telescope, which is operated by the Jet Propulsion Laboratory, California 
%Institute of Technology under a contract with NASA.

The ISO Data Archive is maintained at the ISO Data Centre, Villafranca, 
Madrid, and is part of the Science Operations and Data Systems Division of the Research and 
Scientific Support Department.
ISOPHOT observations were reduced using the ISOPHOT Interactive Analysis package 
PIA, which is a joint development by the ESA Astrophysics Division and the 
ISOPHOT Consortium, lead by the Max-Planck-Institut f\"ur Astronomie (MPIA).

The work was partly supported by the grants OTKA K62304 and T043739 of the 
Hungarian Scientific Research Fund.

\appendix

\section{Bogus debris disks} \label{bogusdisk}

In the last few years several IRAS-based debris disk candidates turned out to be
bogus. Examples are HD\,155826 \citep{lisse}, or the list of \citet{kalas02}. 
Most common problems are contamination by background objects (cirrus knots, 
galaxies), Pleiades like nebulosity, or unreliable point source detection by IRAS. 
Our list of debris disk candidates is based on IRAS data, however when 
higher resolution far-infrared observations were available we checked whether the above mentioned
problems could have affected the detection of the disk.
In Table~\ref{tabbogus} we list those objects which were identified as debris system in the literature (and claimed to have 
$10^{-4} < f_d < 10^{-2}$ in the original paper), but our analysis indicates that they are very likely bogus disks.  
In the following we briefly describe the reason of rejection.

\paragraph{HD\,34739}
The source position in the MIPS 70$\mu$m map differs from the star's position by 26\arcsec, but 
coincides with the near-infrared source 2MASS J05163646-5257397 with an offset of 2\arcsec.

\paragraph{HD\,53842}
This system is not a real bogus disk, since at 24$\mu$m the star shows infrared excess (Mo\'or et al. 2006, in prep.) 
but at 70$\mu$m the IR emission comes from an independent compact source separated by 19\arcsec.
This nearby source coincides with a 2MASS J06460135-8359359, within a distance of 2\arcsec.  
Due to this fact the fractional luminosity of HD\,53842 decreased below our lower limit.

\paragraph{HD\,56099}
The source position in the MIPS 70$\mu$m map differs from the star's position by 24\arcsec, but 
coincides with the near-infrared source 2MASS J07190966+5907219 with an offset of 2\arcsec.

\paragraph{HD\,72390}
The peak brightness position in the ISOPHOT maps at 60 and 90$\mu$m differs from the stellar position by 
36\arcsec.  
Coordinates of this peak are very close to those of 2MASS J08143635-8423260 
which is an extended 2MASS source (XSC 1524951), with an offset of 5\arcsec.

\paragraph{HD\,82821}
We have detected two infrared sources in the MIPS 70$\mu$m map, but their positions significantly 
differ from the position of the star. 
The position of the brighter one lies at 70\arcsec~from the nearby IRAS source (FSC 09319+0346); 
and is located just outside the 2$\sigma$ error ellipse (nearly along the major axis) but well inside the 
3$\sigma$ error ellipse. 
We assume that this source, whose position coincides well with the near-infrared source 2MASS J09343630+0332417 
(XSC 2391850) within a distance of 3\arcsec, is responsible for the source confusion. 
The second source was probably below the IRAS sensitivity limit.
 
\paragraph{HD\,143840}
ISOPHOT mini-map observation at 90$\mu$m and Spitzer MIPS image at 70$\mu$m show extended IR emission. 
The image of the Digitized Sky Survey also shows a reflection nebulosity around this star.
We think that the excess far-infrared emission 
comes from the nebula. 

\paragraph{HD\,185053}
\citet{magakian03} proposed that HD\,185053 is the illuminating source of the reflection nebula GN\,19.41.5.
Spitzer MIPS observation at 24 and 70$\mu$m shows extended IR emission around the star. 
We think that the excess far-infrared emission is related to the nebula rather than to a debris disk.

\paragraph{HD\,204942}
The source position in the MIPS 70$\mu$m map differs from the star's position by 22\arcsec, but 
coincides with the near-infrared source 2MASS J21323602-2409319 with an offset of 4\arcsec.

\paragraph{HD\,23484, HD\,158373, HD\,164330}
ISOPHOT mini-map observations of these stars at 60 and 90$\mu$m did not show excess above the photosphere. 
This discrepancy between the IRAS based excesses and the non-detection of excesses by ISOPHOT (which had better 
spatial resolution than IRAS at far-infrared wavelengths) was already mentioned 
by Silverstone (2000), who suggested cirrus contamination as the reason.

For stars when no higher resolution data were available a set of criteria were applied to filter out 
suspicious objects which might be bogus disk (see Sect.~\ref{rejection}). 
In Table~\ref{tabsusp} we listed those objects which were earlier identified as debris disk in the literature 
(with $10^{-4} < f_d < 10^{-2}$) but were rejected from the further analysis according to our criteria. 
Nevertheless, future high spatial resolution infrared data are needed to prove or disprove our judgment.

\section{New members in the young stellar kinematic groups} \label{newmembers}

\paragraph{$\beta$\,Pictoris moving group (BPMG).}

We identified two stars which are candidate members of this group: HD\,15115 and HD\,191089.
HD\,15115 is a northern object. Though the first surveys found BPMG members only in the 
Southern hemisphere \citep{barrado99,zs01} recently \citet{song03} proposed several new northern candidates. 
One of those, HIP\,12545 is located within four degrees to HD\,15115 on the sky. HD\,15115 has a ROSAT counterpart
with fractional X-ray luminosity of $\rm \log({L_x}/L_{bol})  = -4.94$ ($\rm \log{L_x} = 29.2$\,erg~s$^{-1}$). 
These properties are comparable to those of stars with similar mass in young star associations (see fig.\,9,10 in de la Reza \& Pinz\'on 2004). 

HD\,191089 is somewhat more distant (see Table~\ref{tab1}) than known members of the group in the list of \citet{zs04b}. However,
large part of this list is based on a volume limited survey within $\rm d < 50\,pc$, and other authors proposed candidates at 
larger distances \citep{torres02}.
HD\,191089 shows several signs of youth. \citet{mamajek04b} classified this star as younger than Pleiades because of 
its lithium abundance (EW(Li) = $95\pm 6$\,m$\rm \AA$) was higher than 95\% of Pleiades stars with similar effective temperatures. Isochronal age of the star
($\rm 17^{+8}_{-4}$\,Myr, Mamajek 2004b) is in good agreement with the age of $\rm \beta$\,Pictoris moving group 
($\rm 12^{+8}_{-4}$\,Myr, see Table~\ref{tab2}). Moreover HD\,191089 rotates rapidly compared with typical F5 type stars 
($v\sin{i} = \rm 45$\,km~s$^{-1}$, Nordstr\"om et al. 2004). 
Its ROSAT X-ray luminosity of $\rm \log{L_x} = 29.2$\,erg~s$^{-1}$ and X-ray fractional luminosity of 
$\rm \log({L_x}/L_{bol})  = -4.93$. 
These values are also comparable with the properties of stars with similar spectral types in nearby 
kinematic groups (see fig.\,9,10 in de la Reza \& Pinz\'on 2004).

\paragraph{The GAYA2 and TucHor associations.}

\begin{figure} 
\epsscale{.8}
\plotone{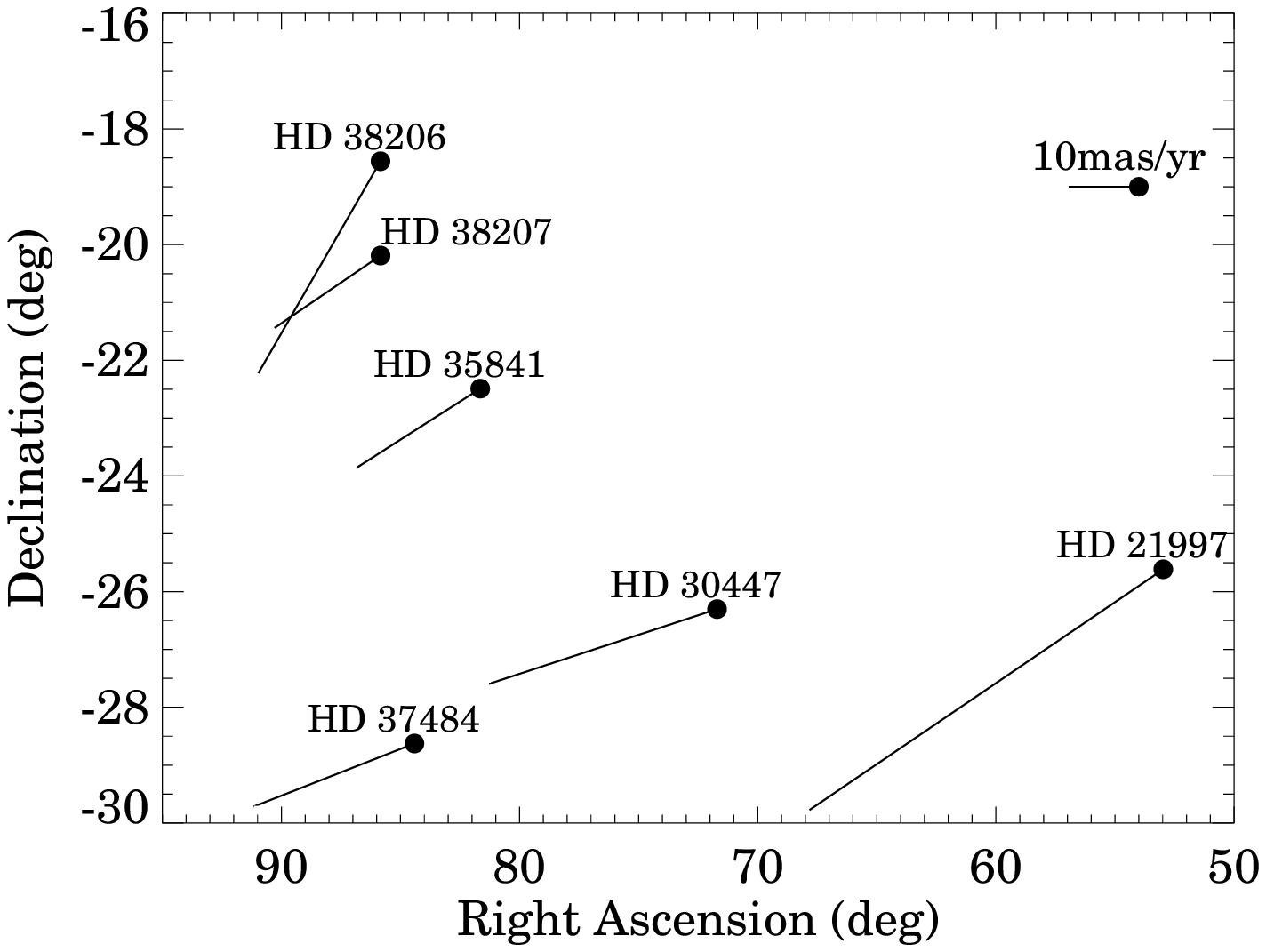}
\caption{Positions and proper motions of proposed new members of the Tucana-Horologium and GAYA2 associations 
(see Appendix\,B.) \label{figgaya}}
\end{figure}

The Great Austral Young Association2 (GAYA2) and Tucana/Horologium (TucHor)
Association can be discussed together since they overlap both in their 
location on the projected sky plane and in velocity space.
GAYA2 was discovered in the framework of the SACY (Search for Associations Containing Young stars) survey \citep{torres02}, 
and the known members are confined mostly in the right ascension (R.A.) range $\rm 3^h < R.A. < 9^h$.
Recently identified members of TucHor occupy a similar region ($ \rm 2^h < R.A. < 7^h$) and several of them
show only slightly different Galactic space motions compared to the mean UVW velocities of the Tucana nucleus and resemble 
the mean space motions of GAYA2. Although GAYA2 is more distant (located at a mean distance of $\rm \sim 84\,pc$, while 
TucHor members located in the same sky region have a mean distance of $\rm \sim 50\,pc$), there is an overlap in radial 
distance, as well.  
Studying the relationship between the two associations is out of the scope of this work. 
As a practical solution, we assigned all doubtful sources (see Fig.~\ref{figgaya}) of $\rm D \leq 67\,pc$ to the TucHor association and the  
more distant ones to GAYA2. 
Thus we propose that one of these sources, HD\,37484 belongs to TucHor (its space velocity is not inconsistent with
that of other neighbouring TucHor members). The measured lithium abundance of the star \citep{favata93} is a strong indication of 
its youth.
HD\,21997, HD\,30447, HD\,35841, HD\,38206 and HD\,38207 are classified as members of the GAYA2 group. It is worth to
mention that these five stars form a spectacular concentration of high $f_d$ debris disks within a relatively 
small area on the sky.

\paragraph{Local Association.}
We propose that HD\,10472, HD\,10638, HD\,218396 and HD\,221853 belong to the Local Association.
HD\,10472 was previously a TucHor candidate \citep{torres00,zs01c}, but recently \citet{zs04b} suggested that its 
membership status is uncertain, thus it may be consistent with our result. 
%thus it does not contradict to our results. 

\begin{figure} 
\epsscale{.9}
\plotone{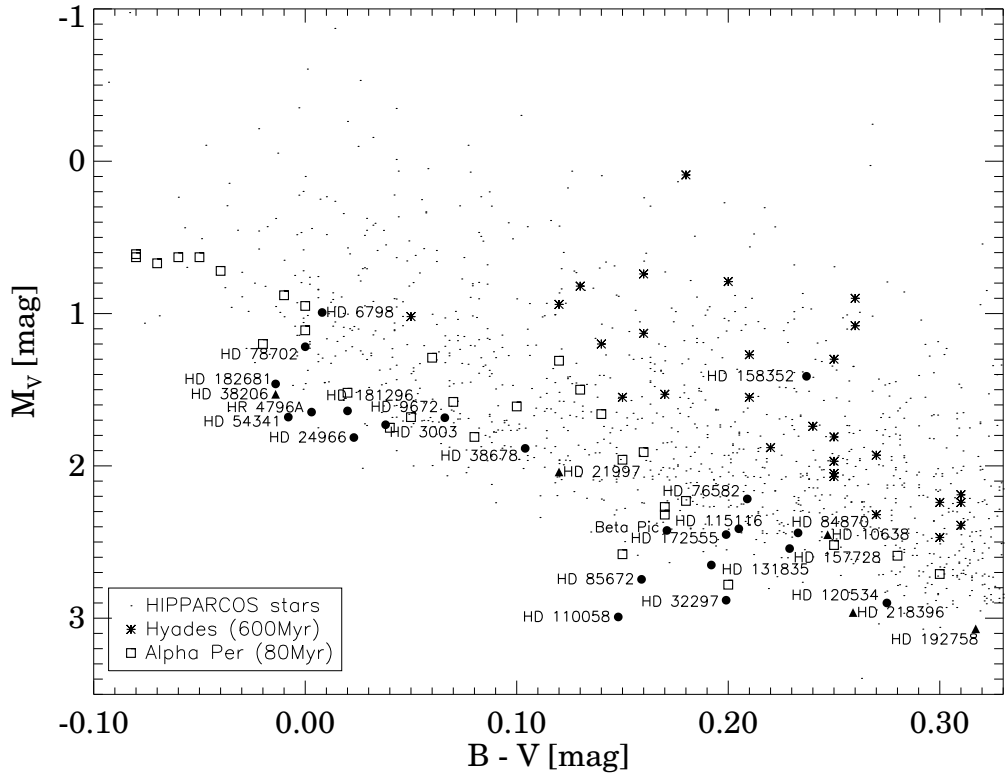}
\caption{Color-magnitude diagram of A-type stars. 
Objects measured by Hipparcos within 100pc from the Sun with parallax error ($<$10\%) 
and (B-V) error ($<$0.01\,mag) are represented 
by small dots. Filled symbols mark positions of A- and F0-type stars from Table~\ref{tab1}; 
triangles correspond to objects assigned to any stellar kinematic group in this work.
Stars of $\alpha$ Per and Hyades open clusters are represented by squares and asterisks, respectively. \label{figcmd}}
\end{figure}

\paragraph{IC 2391 supercluster}
On the basis of its Galactic space velocity HD\,192758 may belong to the IC\,2391 supercluster. Its position on the CMD 
(see Fig.~\ref{figcmd} and Sect.~\ref{cmd}) also suggest its youth.

\paragraph{HD\,110058} was earlier classified as a member of the Lower Centaurus Crux (LCC) association 
using convergent-point method \citep{deZeeuw99}. However, according to our results, its Galactic space velocity 
is inconsistent with the mean velocity of the LCC. Nevertheless HD\,110058 seems to be a very young object on the basis of
its position on the Color-Magnitude Diagram of A-type stars (see Sect.~\ref{cmd}).




\clearpage

\begin{deluxetable}{llcccccccrcc} 
\tabletypesize{\scriptsize}
\rotate
\tablecaption{Stellar properties and derived fractional luminosities of the disks \label{tab1}}
\tablewidth{0pt}
\tablehead{
\colhead{Name} & \colhead{Other} & \colhead{Spectral} & \colhead{V} &
\colhead{B-V} & \colhead{Distance} & \colhead{First reference} & \colhead{Debris disk} 
& \colhead{Reference} & \colhead{$\rm f_d$} & \colhead{Age} & \colhead{Age} \\
\colhead{} & \colhead{Name} & \colhead{Type} & \colhead{} &
\colhead{} & \colhead{} & \colhead{to debris disk} & \colhead{confirmation} 
& \colhead{} & \colhead{} & \colhead{} & \colhead{Reference} \\
\colhead{} & \colhead{} & \colhead{} & \colhead{(mag)} &
\colhead{(mag)} & \colhead{(pc)} & \colhead{} & \colhead{} 
& \colhead{} & \colhead{($\rm 10^{-4}$)} & \colhead{(Myr)} & \colhead{}
}
\startdata
    HD 105  &  ...   &        G0V & 7.51 &  0.595 &  40      & (27)       & ISO/MIPS &  (9,27)	    &	2.5$\pm$0.3 &  $\rm 30^{+10}_{-20} $	 &  Table\,4  \\  
    HD 377  &  ...   &        G2V & 7.59 &  0.626 &  40      &  (9)       & MIPS     &  (9)          &   4.0$\pm$0.3 &     [30,100]	    &  (4,30)  \\
   HD 3003  &  ...   &        A0V & 5.07 &  0.038 &  46      & (29)	  & MIPS$^{*}$     &  ...	    &   1.4$\pm$0.2 & $\rm 30^{+10}_{-20} $      &  Table\,4  \\
   HD 6798  &  ...   &        A3V & 5.60 &  0.008 &  83      & (20)	  & IRS     &  (12)         &   1.6$\pm$0.3 & $\rm 340^{+60}_{-80} $	  &  this work  \\
   HD 8907  &  ...   &         F8 & 6.66 &  0.505 &  34      & (27)	  & ISO/MIPS &  (9,27)        &   2.4$\pm$0.1 &    [100,870]	       &  (4,35,36)  \\
   HD 9672  & 49\,Cet&        A1V & 5.62 &  0.066 &  61      & (24)       & ISO     &  (33)          &   9.2$\pm$0.6 &        [8,20]		       &  (32,36)  \\
  HD 10472  &  ...   &     F2IV/V & 7.62 &  0.420 &  67      & (27)	  & MIPS$^{*}$     &  (21)         &   3.4$\pm$0.9 &	[20,150]	       &  Table\,4  \\
  HD 10647  &  ...   &        F8V & 5.52 &  0.551 &  17      & (31)	  & ISO/MIPS$^{*}$ &  (6)     	    &   3.0$\pm$0.3 & [300,7000] &  (18,36)  \\
  HD 10638  &  ...   &         A3 & 6.73 &  0.247 &  72      & (27)	  & ...	    &  ...          &   3.9$\pm$0.5 &	     [20,150]	       &  Table\,4  \\
  HD 15115  &  ...   &         F2 & 6.79 &  0.399 &  45      & (27)	  & ISO/MIPS &  (27)	    &	4.9$\pm$0.4 & $\rm 12^{+8}_{-4}$	 &  Table\,4  \\
  HD 15745  &  ...   &         F0 & 7.47 &  0.360 &  64      & (27)	  & ISO/MIPS &  (27)	    &  20.1$\pm$1.4 &	   $<$700		 &  (18,36)  \\
  HD 16743  &  ...   &F0/F2III/IV & 6.78 &  0.387 &  60      & this work  & MIPS     &  ...	    &	3.6$\pm$0.3 &$\rm 1200^{+500}_{-600}$	 &  (18)  \\
  HD 17390  &  ...   &     F3IV/V & 6.48 &  0.387 &  45      & (27)       & ISO/MIPS$^{*}$ &  (1)	  &   1.9$\pm$0.2 &	    $<$800	       &  (18,36)  \\
  HD 21997  &  ...   &     A3IV/V & 6.38 &  0.120 &  74      & (20)	  & IRS     &  (12)       &   4.7$\pm$0.3 &  $\rm 20^{+10}_{-10}$	       &  Table\,4  \\
  HD 24966  &  ...   &        A0V & 6.89 &  0.023 & 104      & (27)	  & ...	    &   ... 	  &   2.4$\pm$0.5 &		   $<$100      &  this work  \\
  HD 25457  &  ...   &        F5V & 5.38 &  0.516 &  19      & (27)	  & ISO/MIPS &  (9,27)      &   1.0$\pm$0.2 & [50,100]		       &  Table\,4   \\
  HD 30447  &  ...   &        F3V & 7.85 &  0.393 &  78      & (27)	  & ISO/MIPS &  (1)	  &   7.5$\pm$1.1 & $\rm 20^{+10}_{-10}$       &  Table\,4  \\
  HD 32297  &  ...   &         A0 & 8.13 &  0.199 & 112      & (27)	  & COR     &  (26)    &  33.4$\pm$6.0 &	   $<$30	       &  (13)  \\
  HD 35841  &  ...   &        F5V & 8.91 &  0.496 & $\rm 104^P$ & (27)    & ...	    &  ...       &  15.5$\pm$3.7 &     $\rm 20^{+10}_{-10}$   &  Table\,4  \\
  HD 37484  &  ...   &        F3V & 7.26 &  0.404 &  60      & (20)	  & ISO/MIPS &  (9,27)	  &   2.7$\pm$0.5 & $\rm 30^{+10}_{-20}$       &  Table\,4  \\
  HD 38207  &  ...   &        F2V & 8.47 &  0.391 & $\rm 103^P$ & (27)    & ISO/MIPS &  (9,27)&  10.8$\pm$0.6 & $\rm 20^{+10}_{-10}$	       &  Table\,4  \\
  HD 38206  &  ...   &        A0V & 5.73 & -0.014 &  69      & (17)	  & MIPS     &  (22)	  &   1.4$\pm$0.3 & $\rm 20^{+10}_{-10}$       &  Table\,4 \\
  HD 38678  & $\rm \zeta\,Lep$ & A2Vann & 3.55 &  0.104 &  22 	& (19)    & ISO/MIPS$^{*}$ &  (7)  &   1.1$\pm$0.2 & $\rm 200^{+100}_{-100}$    &  Table\,4  \\
  HD 39060  & $\rm \beta\,Pic$  &  A3V & 3.85 &  0.171 &  19 	& (28)    & ISO/COR/SUBM&  (8,10,28)&  24.3$\pm$1.1 & $\rm 12^{+8}_{-4}$         &  Table\,4  \\
  HD 50571  &  ...   &   F7III-IV & 6.11 &  0.457 &  33      &  this work & MIPS$^{*}$ &  ...	  &   1.1$\pm$0.3 & $\rm 1800^{+1000}_{-1300}$ &  (18)  \\
  HD 53143  &  ...   &     K0IV-V & 6.81 &  0.786 &  18      & (17)	  & ISO     &  (6)	  &   2.0$\pm$0.5 & $\rm 980^{+520}_{-330}$    &  (6)  \\
  HD 54341  &  ...   &        A0V & 6.52 & -0.008 &  93      &  this work & ...	    &  ...	  &   1.8$\pm$0.4 &   $<$100		       &  this work  \\
  HD 69830  &  ...   &        K0V & 5.95 &  0.754 &  13      & (17)	  & IRS     &  (3)	  &   2.0\tablenotemark{a}&  [600,4700] 	       &  (29,35)  \\
  HD 76582  &  ...   &       F0IV & 5.68 &  0.209 &  49      & (27)	  & ...	    &  ...	  &   1.7$\pm$0.2 & $\rm 450^{+150}_{-290}$    &  this work  \\
  HD 78702  &  ...   &     A0/A1V & 5.73 &  0.000 &  80      & (27)	  & ...	    &  ...	  &   2.1$\pm$0.7 &   $\rm 220^{+100}_{-140}$  &  this work  \\
  HD 84870A &  ...   &         A3 & 7.20 &  0.233 &  90      & (27)	  & ...	    &  ...	  &   4.5$\pm$1.0 &	      $<$520	      &  this work  \\
  HD 85672  &  ...   &         A0 & 7.59 &  0.159 &  93      & (27)	  & ...	    &  ...	  &   4.9$\pm$1.0 &	      $<$100	      &  this work  \\
  HD 92945  &  ...   &        K1V & 7.72 &  0.873 &  22      & (27)	  & MIPS     &  (5)	  &   5.3$\pm$1.2 & 100                       &  (30,35)   \\
 HD 107146  &  ...   &        G2V & 7.04 &  0.604 &  29      & (27)	  & MIPS/COR/SUBM & (2,9,34)      &   9.2$\pm$0.9 & $\rm  100^{+100}_{-20}$    &  (34,36)  \\
 HD 109573A & HR\,4796A  &    A0V & 5.78 &  0.003 &  67      & (11)       & MIR/COR &(14,25)   &  47.0$\pm$2.5 & $\rm  8^{+7}_{-3}$         &  Table\,4  \\
 HD 110058  &  ...   &        A0V & 7.99 &  0.148 & 100      & (17)	  & ISO     &  (1)	  &  18.9$\pm$3.3 &	  $<$100	       &  this work  \\
 HD 115116  &  ...   &        A7V & 7.07 &  0.205 &  85      &  this work & ...	    &  ...	 &   3.2$\pm$1.0 &	       $<$360	      &  this work  \\
 HD 120534  &  ...   &        A5V & 7.02 &  0.275 & $\rm 67^P$  & (27)    & ...     &  ...	 &   4.9$\pm$0.9 &	       $<$320	      &  this work  \\
 HD 121812  &  ...   &         K0 & 8.53 &  0.820 &  38      &  this work & ...	    &  ...	  &  14.9$\pm$4.1 &    $\rm 230^{+150}_{-90}$  &  this work  \\
 HD 122106  &  ...   &        F8V & 6.36 &  0.486 &  78      &  this work & ...	    &  ...	  &   1.2$\pm$0.4 &  [1000,1600]	       &  (18,23)   \\
 HD 127821  &  ...   &       F4IV & 6.10 &  0.428 &  32      & (27)	  & MIPS$^{*}$&  ...	  &   2.2$\pm$0.4 &  [200,3400] 	       &  (18,36)  \\
 HD 130693  &  ...   &        G6V & 8.20 &  0.734 & $\rm 33^P$  &  this work & ...  &  ...       &   5.8$\pm$1.9 &    $\rm  50^{+50}_{-40}$   &  this work  \\
 HD 131835  &  ...   &       A2IV & 7.88 &  0.192 & 111      &  this work & ...	 & ...        &  19.9$\pm$3.3 &    $\rm  17^{+1}_{-1}$     &  Table\,4  \\
 HD 157728  &  ...   &       F0IV & 5.70 &  0.229 &  43      & (31)	  & ...	 & ...        &   2.9$\pm$0.4 & $<$200  		   &  this work  \\
 HD 158352  &  ...   &        A8V & 5.41 &  0.237 &  63      & (19)	  & ...	 & ...        &   1.8$\pm$0.5 &   $\rm 750^{+150}_{-150}$  &  this work  \\
 HD 164249  &  ...   &        F5V & 7.01 &  0.458 &  47      & (27)	  & ISO/MIPS$^{*}$ &  (27)	  &  10.4$\pm$1.6 & $\rm  12^{+8}_{-4}$        &  Table\,4  \\
 HD 169666  &  ...   &         F5 & 6.68 &  0.444 &  51      &  this work & MIPS     &  ...	  &   1.5$\pm$0.3 & $\rm 2200^{+400}_{-600}$   &  (18)  \\
 HD 170773  &  ...   &        F5V & 6.22 &  0.429 &  36      & (24)	  & ISO/MIPS$^{*}$ &  (27)	  &   3.8$\pm$0.4 &   [200,2800]	       &  (18,36)  \\
 HD 172555  &  ...   &        A7V & 4.78 &  0.199 &  29      & (19)	  & MIPS$^{*}$     &  ...	  &   7.8$\pm$0.7 & $\rm  12^{+8}_{-4}$        &  Table\,4  \\
 HD 181296  &  $\rm \eta\,Tel$      &   A0Vn & 5.03 &  0.020 &  48 & (17) & ISO/MIPS$^{*}$ &  (27)	  &   2.4$\pm$0.2 & $\rm  12^{+8}_{-4}$        &  Table\,4  \\
 HD 181327  &  ...   &     F5/F6V & 7.04 &  0.480 &  51      & (17)	  & ISO/MIPS$^{*}$ &  (1)	  &  29.3$\pm$1.6 & $\rm  12^{+8}_{-4}$        &  Table\,4  \\
 HD 182681  &  ...   &     B8/B9V & 5.66 & -0.014 &  69      &  this work & ...	    &  ...	  &   1.5$\pm$0.3 & $<$100		       &  this work  \\
 HD 191089  &  ...   &        F5V & 7.18 &  0.480 &  54      & (17)	  & ISO/MIPS$^{*}$ &  (1)	  &  19.1$\pm$2.2 & $\rm  12^{+8}_{-4}$        &  Table\,4  \\
 HD 192758  &  ...   &        F0V & 7.02 &  0.317 & $\rm 62^P$  & (27)    & ISO/MIPS &  (1)  &   5.6$\pm$0.5 &     [35,55]		       &  Table\,4  \\
 HD 197481  & AU\,Mic&       M1Ve & 8.81 &  1.470 &  10      & (27)	  & MIPS/COR/SUBM & (5,15,16)&   4.0$\pm$0.3 & $\rm  12^{+8}_{-4}$        &  Table\,4  \\
 HD 202917  &  ...   &        G5V & 8.65 &  0.690 &  46      & (27)	  & MIPS$^{*}$  &  (27)	  &   2.9$\pm$0.8 & $\rm  30^{+10}_{-20}$      &  Table\,4  \\
 HD 205674  &  ...   &    F3/F5IV & 7.19 &  0.396 &  53      &  this work & MIPS$^{*}$	&  ...	  &   3.5$\pm$0.8 &  $\rm 2600^{+900}_{-1400}$ &  (18)  \\
 HD 206893  &  ...   &        F5V & 6.69 &  0.439 &  39      & (27)	  & ISO/MIPS$^{*}$ &  (27)	  &   2.3$\pm$0.2 &	   $<$2800	       &  (18,36)  \\
 HD 218396  &  ...   &        A5V & 5.97 &  0.259 &  40      & (27)	  & ISO           &  (27)	  &   2.3$\pm$0.2 &	   [20,150]	       &  Table\,4  \\
 HD 221853  &  ...   &         F0 & 7.35 &  0.405 &  71      & (27)	  & ISO/MIPS &  (27)	  &   8.0$\pm$1.1 &	  [20,150]	       & Table\,4 \\
 \enddata
 %% Text for table notes should follow after the \enddata but before
 %% the \end{deluxetable}. Make sure there is at least one \tablenotemark
 %% in the table for each \tablenotetext.
 \tablecomments{Col.(1): Names. Col.(2): Other names. 
 Cols.(3-5): Data are from the Hipparcos or the Tycho-2 Spectral Type Catalog.
 Col.(6): Distances. P indicates photometric distances otherwise Hipparcos 
 distances were used. Col.(7): Reference for first identification as debris disks.  Col.(8): Observations following 
 the original IRAS discovery which independently confirmed the existence of the debris disk.   
 {\it COR}: Coronographic observation; {\it ISO}: ISOPHOT; {\it MIR}: Observation at mid-infrared wavelengths; 
 {\it IRS}: Spitzer IRS; {\it MIPS}: Spitzer MIPS; {\it SUBM}: Observation at submillimeter wavelengths.
 $^{*}$ marks those MIPS observations when we extracted only astrometrical information from the maps, but 
 did not determine photometric fluxes.
 Col.(9): References for papers related to observations of Col.(8).  
 Col.(10): Fractional dust luminosities and their uncertainties. 
 Col.(11): Ages estimates. [t1,t2] marks an age interval. Col.(12) References for age estimates.}
\tablenotetext{a}{Beichman et al. (2005b) demonstrated the presence of strong spectral features for this object, therefore we took their 
fractional luminosity estimate rather than fitting the photometric points by a modified blackbody} 

\tablerefs{(1) \'Abrah\'am et al., 2006, in prep.; (2) Ardila et al. (2004); (3) Beichman et al. (2005b); (4) Carpenter et al. (2005); 
(5) Chen et al. (2005b); 
 (6) Decin et al. (2000); (7) Habing et al. (2001); (8) Heinrichsen et al.(1999);
 (9) {\sl http://data.spitzer.caltech.edu/popular/feps/20051123\_enhanced\_v1/}, 
 FEPS Data Explanatory Supplement v3.0, Hines et al. (2005); (10) Holland et al. (1998);  
 (11) Jura (1991); (12) Jura et al. (2004); (13) Kalas (2005); 
  (14) Koerner et al. (1998); (15) Krist et al. (2005); (16) Liu et al. (2004); 
 (17) Mannings \& Barlow (1998);  (18) Nordstr\"om et al. (2004); 
 (19) Oudmaijer et al. (1992); (20) Patten \& Willson (1991); (21) Rebull et al. (2005); (22) Rieke et al. (2005); 
 (23) Rocha-Pinto et al. (2004); 
 (24) Sadakane \& Nishida (1986);
 (25) Schneider et al. (1999); (26) Schneider et al. (2005); (27) Silverstone (2000); (28) Smith \& Terrile (1984); 
 (29) Song et al. (2000); (30) Song et al. (2004); (31) Stencel \& Backman (1991); (32) Thi et al.(2001); 
 (33) Walker and Heinrichsen (2000); 
 (34) Williams et al. (2004);
 (35) Wright et al. (2004); 
 (36) Zuckerman \& Song (2004a); 'Table\,4' refers to ages of stellar kinematic groups, as presented in Table\,4.}

 \end{deluxetable}

%%%%%%%%%%%%%%%%%%%%%%%%%%%%%%%%%%%%%%%%%%%%%%%%%%%%%%%%%%%%%%%%%%%%%%%%%%%%%%%%%%%%%%%%%%%%%%%%%%%%%%%%%%%%%%%%%%%%%%
%%%%%%%%%%%%%%%%%%%%%%%%%%%%%%%%%%%%%%%%%%%%%%%%%%%%%%%%%%%%%%%%%%%%%%%%%%%%%%%%%%%%%%%%%%%%%%%%%%%%%%%%%%%%%%%%%%%%%%

\begin{deluxetable}{llcccccc} 
\tabletypesize{\scriptsize}
\tablecaption{Summary table of fluxes measured in excess \label{tabflux}}
\tablehead{
\colhead{Name} & \colhead{Instrument} & \colhead{Wavelength} & \multicolumn{2}{c}{Measured flux} & 
\colhead{Flux} & \colhead{Photosph.} & \colhead{Ref.} \\
\colhead{} & \colhead{} & \colhead{} & \colhead{uncorrected} & 
\colhead{corrected} & \colhead{quality} & \colhead{flux} & \colhead{} \\
\colhead{} & \colhead{} & \colhead{[$\mu$m]} & \colhead{[Jy]} & 
\colhead{[Jy]} & \colhead{} & \colhead{[Jy]} & \colhead{}
}
\startdata
      HD 105 & ISOPHOT &   60 &   0.107$\pm$0.014 &   0.108$\pm$0.014 & ... &	 3.9e-03 & 1 \\
             &    MIPS &   70 &   0.145$\pm$0.012 &   0.164$\pm$0.014 & ... &	 2.9e-03 & 4 \\
             & ISOPHOT &   90 &   0.147$\pm$0.010 &   0.159$\pm$0.011 & ... &	 1.7e-03 & 1 \\
             &    MIPS &  160 &   0.166$\pm$0.028 &   0.171$\pm$0.029 & ... &	 5.5e-04 & 4 \\
\hline
      HD 377 &    MIPS &   24 &   0.035$\pm$0.002 &   0.035$\pm$0.002 & ... &      0.025 & 4 \\
             &    IRAS &   60 &   0.179$\pm$0.014 &   0.185$\pm$0.015 &   3 &	 4.0e-03 & 5 \\
             &    MIPS &   70 &   0.146$\pm$0.017 &   0.158$\pm$0.018 & ... &	 2.9e-03 & 4 \\
\hline
     HD 3003 &    IRAS &   25 &   0.343$\pm$0.027 &   0.320$\pm$0.026 &   3 &	0.063	 & 8 \\
\hline
     HD 6798 &    IRAS &   25 &   0.111$\pm$0.018 &   0.108$\pm$0.017 &   3 &	0.039	 & 8 \\
             &    IRAS &   60 &   0.403$\pm$0.048 &   0.414$\pm$0.050 &   2 &	 6.7e-03 & 8 \\
\hline
     HD 8907 &    IRAS &   60 &   0.285$\pm$0.054 &   0.305$\pm$0.058 &   3 &	 7.4e-03 & 8 \\
             & ISOPHOT &   60 &   0.224$\pm$0.020 &   0.232$\pm$0.021 & ... &	 7.4e-03 & 1 \\
             &    MIPS &   70 &   0.232$\pm$0.007 &   0.256$\pm$0.008 & ... &	 5.4e-03 & 4 \\
             & ISOPHOT &   90 &   0.264$\pm$0.018 &   0.270$\pm$0.018 & ... &	 3.3e-03 & 1 \\
\hline
     HD 9672 &    IRAS &   25 &   0.384$\pm$0.042 &   0.438$\pm$0.048 &   3 &	   0.041 & 8 \\
             &    IRAS &   60 &    2.02$\pm$0.12  &    2.09$\pm$0.13  &   3 &	 7.0e-03 & 8 \\
             &    IRAS &  100 &    1.88$\pm$0.21  &    1.85$\pm$0.20  &   2 &	 2.5e-03 & 8 \\
             & ISOPHOT &  120 &    1.45$\pm$0.38  &    1.32$\pm$0.35  & ... &	 1.7e-03 & 1 \\
             & ISOPHOT &  170 &   0.862$\pm$0.181 &   0.772$\pm$0.162 & ... &	 8.7e-04 & 1 \\
\hline
    HD 10472 &    IRAS &   60 &   0.140$\pm$0.032 &   0.145$\pm$0.033 &   2 &	 2.4e-03 & 8 \\
\hline
    HD 10647 &    IRAS &   60 &   0.815$\pm$0.106 &   0.877$\pm$0.114 &   3 &	   0.024 & 8 \\
             & ISOPHOT &   60 &   0.803$\pm$0.040 &   0.824$\pm$0.041 & ... &	   0.024 & 1 \\
             &    IRAS &  100 &    1.08$\pm$0.17  &    1.08$\pm$0.17  &   2 &	 8.7e-03 & 8 \\
\hline
    HD 10638 &    IRAS &   60 &   0.348$\pm$0.038 &   0.360$\pm$0.040 &   3 &	 3.6e-03 & 8 \\
\hline
    HD 15115 &    MIPS &   24 &   0.055$\pm$0.006 &   0.054$\pm$0.006 & ... &    0.032   & 7 \\
             &    IRAS &   60 &   0.441$\pm$0.048 &   0.473$\pm$0.052 &   3 &	 5.1e-03 & 8 \\
             & ISOPHOT &   60 &   0.401$\pm$0.019 &   0.415$\pm$0.020 & ... &	 5.1e-03 & 1 \\
             & ISOPHOT &   90 &   0.419$\pm$0.029 &   0.427$\pm$0.030 & ... &	 2.2e-03 & 1 \\
\hline
    HD 15745 &    MIPS &   24 &   0.162$\pm$0.016 &   0.168$\pm$0.017 & ... &	   0.016 & 7 \\
             &    IRAS &   25 &   0.177$\pm$0.028 &   0.207$\pm$0.033 &   2 &	   0.015 & 8 \\
             &    IRAS &   60 &   0.868$\pm$0.061 &   0.875$\pm$0.061 &   3 &	 2.6e-03 & 8 \\
             &     ISO &   60 &   0.753$\pm$0.038 &   0.770$\pm$0.039 & ... &	 2.6e-03 & 1 \\
             &     ISO &   90 &   0.532$\pm$0.037 &   0.515$\pm$0.036 & ... &	 1.1e-03 & 1 \\
\hline
    HD 16743 &    MIPS &   24 &   0.049$\pm$0.005 &   0.049$\pm$0.005 & ... &	   0.029 & 7 \\
             &    IRAS &   60 &   0.304$\pm$0.033 &   0.323$\pm$0.036 &   3 &	 4.6e-03 & 8 \\
\hline
    HD 17390 &    IRAS &   60 &   0.226$\pm$0.034 &   0.240$\pm$0.036 &   3 &	 6.4e-03 & 8 \\
             &     ISO &   60 &   0.253$\pm$0.021 &   0.262$\pm$0.022 & ... &	 6.4e-03 & 1 \\
             &     ISO &   90 &   0.228$\pm$0.023 &   0.231$\pm$0.023 & ... &	 2.8e-03 & 1 \\
\hline
    HD 21997 &    IRAS &   60 &   0.595$\pm$0.036 &   0.646$\pm$0.039 &   3 &	 3.9e-03 & 8 \\
             &    IRAS &  100 &   0.636$\pm$0.108 &   0.629$\pm$0.107 &   2 &	 1.4e-03 & 8 \\
\hline
    HD 24966 &    IRAS &   60 &   0.184$\pm$0.031 &   0.190$\pm$0.032 &   3 &	 1.9e-03 & 8 \\
\hline
    HD 25457 &    MIPS &   24 &   0.202$\pm$0.011 &   0.204$\pm$0.011 & ... &	   0.147 & 4 \\
             &    IRAS &   60 &   0.287$\pm$0.075 &   0.285$\pm$0.074 &   2 &	   0.023 & 8 \\
             & ISOPHOT &   60 &   0.293$\pm$0.030 &   0.297$\pm$0.030 & ... &	   0.023 & 1 \\
             &    MIPS &   70 &   0.288$\pm$0.012 &   0.310$\pm$0.013 & ... &	   0.017 & 4 \\
             & ISOPHOT &   90 &   0.246$\pm$0.018 &   0.242$\pm$0.018 & ... &	   0.010 & 1 \\
\hline
    HD 30447 &    MIPS &   24 &   0.028$\pm$0.003 &   0.028$\pm$0.003 & ... &	   0.012 & 7 \\
             &    IRAS &   60 &   0.274$\pm$0.030 &   0.296$\pm$0.033 &   3 &	 1.9e-03 & 8 \\
             &     ISO &   60 &   0.247$\pm$0.062 &   0.256$\pm$0.064 & ... &	 1.9e-03 & 1 \\
             &     ISO &   90 &   0.271$\pm$0.065 &   0.277$\pm$0.067 & ... &	 8.3e-04 & 1 \\
\hline
    HD 32297 &    IRAS &   25 &   0.211$\pm$0.034 &   0.256$\pm$0.041 &   2 &	 5.7e-03 & 8 \\
             &    IRAS &   60 &    1.12$\pm$0.07  &    1.14$\pm$0.07  &   3 &	 9.7e-04 & 8 \\
\hline
    HD 35841 &    IRAS &   60 &   0.188$\pm$0.040 &   0.195$\pm$0.041 & ... &	 8.1e-04 & 8 \\
\hline
    HD 37484 &    MIPS &   24 &   0.053$\pm$0.003 &   0.055$\pm$0.003 & ... &     0.021  & 4 \\
             &    IRAS &   60 &   0.128$\pm$0.036 &   0.124$\pm$0.035 &   2 &	 3.3e-03 & 8 \\
             & ISOPHOT &   60 &   0.094$\pm$0.012 &   0.095$\pm$0.012 & ... &	 3.3e-03 & 1 \\
	     &    MIPS &   70 &   0.104$\pm$0.011 &   0.110$\pm$0.012 & ... &    2.4e-03 & 4 \\
             & ISOPHOT &   90 &   0.047$\pm$0.012 &   0.044$\pm$0.011 & ... &	 1.5e-03 & 1 \\
\hline
    HD 38207 &    MIPS &   24 &   0.017$\pm$0.003 &   0.017$\pm$0.003 & ... &	 6.9e-03 & 4 \\
             & ISOPHOT &   60 &   0.205$\pm$0.015 &   0.212$\pm$0.015 & ... &	 1.1e-03 & 1 \\
             &    MIPS &   70 &   0.178$\pm$0.006 &   0.194$\pm$0.007 & ... &	 7.9e-04 & 4 \\
             & ISOPHOT &   90 &   0.177$\pm$0.013 &   0.177$\pm$0.013 & ... &	 4.8e-04 & 1 \\
\hline
    HD 38206 &    MIPS &   24 &   0.115$\pm$0.023 &   0.119$\pm$0.024 & ... &	 0.033   & 9 \\  
             &    IRAS &   25 &   0.110$\pm$0.021 &   0.113$\pm$0.021 &   2 &	 0.030   & 8 \\
             &    IRAS &   60 &   0.313$\pm$0.038 &   0.313$\pm$0.038 &   3 &	 5.1e-03 & 8 \\
\hline
    HD 38678 &    IRAS &   25 &    1.16$\pm$0.09  &    0.97$\pm$0.08 &   3 &	   0.296 & 8 \\
             &    IRAS &   60 &   0.366$\pm$0.073 &   0.302$\pm$0.060 &   3 &	   0.051 & 8 \\
             & ISOPHOT &   60 &   0.404$\pm$0.020 &   0.390$\pm$0.019 & ... &	   0.051 & 1 \\
\hline
    HD 39060 &    IRAS &   12 &    3.40$\pm$0.14  &    2.38$\pm$0.10  &   3 &	   1.09  & 8 \\
             &    IRAS &   25 &    8.81$\pm$0.35  &   10.29$\pm$0.41  &   3 &	   0.254 & 8 \\
             &    IRAS &   60 &    19.7$\pm$0.79  &   18.92$\pm$0.76  &   3 &	   0.043 & 8 \\
             &    IRAS &  100 &    11.0$\pm$0.44  &    9.32$\pm$0.37  &   3 &	   0.015 & 8 \\
             & ISOPHOT &  120 &     7.9$\pm$1.2   &    6.68$\pm$1.01  & ... &	   0.011 & 1 \\
             & ISOPHOT &  150 &    4.80$\pm$0.73  &    4.39$\pm$0.67  & ... &	 6.8e-03 & 1 \\
             & ISOPHOT &  170 &     4.0$\pm$0.6   &    3.29$\pm$0.49  & ... &	 5.4e-03 & 1 \\
             & ISOPHOT &  200 &    2.10$\pm$0.35  &    2.03$\pm$0.34  & ... &	 3.9e-03 & 1 \\
\hline
    HD 50571 &    IRAS &   60 &   0.183$\pm$0.040 &   0.186$\pm$0.041 &   2 &	   0.011 & 8 \\
\hline
    HD 53143 &    IRAS &   60 &   0.152$\pm$0.040 &   0.155$\pm$0.040 &   2 &	   0.012 & 8 \\
             & ISOPHOT &   60 &   0.214$\pm$0.011 &   0.219$\pm$0.011 & ... &	   0.012 & 1 \\
             & ISOPHOT &   90 &   0.145$\pm$0.011 &   0.142$\pm$0.011 & ... &	 5.1e-03 & 1 \\
\hline
    HD 54341 &    IRAS &   60 &   0.203$\pm$0.047 &   0.210$\pm$0.048 &   2 &	 2.7e-03 & 8 \\
\hline
    HD 69830 &    IRAS &   25 &   0.341$\pm$0.038 &   0.292$\pm$0.032 &   3 &	   0.142 & 8 \\
\hline
    HD 76582 &    IRAS &   60 &   0.391$\pm$0.047 &   0.403$\pm$0.048 &   3 &	 9.2e-03 & 8 \\
\hline
    HD 78702 &    IRAS &   60 &   0.314$\pm$0.047 &   0.327$\pm$0.049 &   2 &	 5.6e-03 & 8 \\
             &    IRAS &  100 &   0.954$\pm$0.267 &   0.987$\pm$0.276 &   2 &	 2.0e-03 & 8 \\
\hline
    HD 84870A&    IRAS &   60 &   0.243$\pm$0.046 &   0.252$\pm$0.048 &   3 &	 2.4e-03 & 8 \\
\hline
    HD 85672 &    IRAS &   60 &   0.193$\pm$0.037 &   0.200$\pm$0.038 &   3 &	 1.4e-03 & 8 \\
\hline
    HD 92945 &    IRAS &   60 &   0.248$\pm$0.045 &   0.270$\pm$0.049 &   3 &	 6.3e-03 & 8 \\
             &    MIPS &   70 &   0.271$\pm$0.060 &   0.301$\pm$0.067 & ... &	 4.6e-03 & 2 \\
\hline
   HD 107146 &    MIPS &   24 &   0.060$\pm$0.003 &   0.059$\pm$0.003 & ... &      0.042 & 4 \\
             &    IRAS &   60 &   0.705$\pm$0.056 &   0.777$\pm$0.062 &   3 &	 6.7e-03 & 8 \\
	     &    MIPS &   70 &   0.648$\pm$0.014 &   0.727$\pm$0.016 & ... &    4.9e-03 & 4 \\
             &    IRAS &  100 &   0.910$\pm$0.155 &   0.913$\pm$0.155 &   2 &	 2.4e-03 & 8 \\
	     &   SCUBA &  450 &   0.130$\pm$0.040 &   0.130$\pm$0.040 &  ...&    1.1e-04 & 10\\
	     &   SCUBA &  850 &   0.020$\pm$0.004 &   0.020$\pm$0.004 &  ...&    3.2e-05 & 10\\
\hline
   HD 109573A&    IRAS &   12 &   0.284$\pm$0.011 &   0.214$\pm$0.009 &   3 &	   0.131 & 5 \\
             &    IRAS &   25 &    3.73$\pm$0.00\tablenotemark{a}  &    4.38$\pm$0.00\tablenotemark{a}   &   3 &	  0.031 & 5 \\
             &    IRAS &   60 &    8.07$\pm$0.00\tablenotemark{a}  &    7.71$\pm$0.00\tablenotemark{a}   &   3 &	5.2e-03 & 5 \\
             &    IRAS &  100 &    4.09$\pm$0.08 &    3.86$\pm$0.08  &   3 &	 1.9e-03 & 5 \\
             &   SCUBA &  850 &   0.019$\pm$0.004 &   0.019$\pm$0.004 & ... &	 2.5e-05 & 3  \\
\hline
   HD 110058 &    IRAS &   25 &   0.266$\pm$0.029 &   0.300$\pm$0.033 &   3 &	 5.7e-03 & 8 \\
             &    IRAS &   60 &   0.368$\pm$0.062 &   0.341$\pm$0.058 &   2 &	 9.8e-04 & 8 \\
             & ISOPHOT &   60 &   0.387$\pm$0.085 &   0.387$\pm$0.085 & ... &	 9.8e-04 & 1 \\
             & ISOPHOT &   90 &   0.218$\pm$0.050 &   0.201$\pm$0.046 & ... &	 4.3e-04 & 1 \\
\hline
   HD 115116 &    IRAS &   60 &   0.192$\pm$0.054 &   0.198$\pm$0.056 &   2 &	 2.5e-03 & 8 \\
\hline
   HD 120534 &    IRAS &   60 &   0.315$\pm$0.047 &   0.326$\pm$0.049 & ... &	 3.3e-03 & 8 \\
\hline
   HD 121812 &    IRAS &   60 &   0.263$\pm$0.047 &   0.289$\pm$0.052 &   3 &	 2.7e-03 & 8 \\
             &    IRAS &  100 &   0.484$\pm$0.135 &   0.492$\pm$0.138 &   2 &	 9.6e-04 & 8 \\
\hline
   HD 122106 &    IRAS &   60 &   0.161$\pm$0.042 &   0.164$\pm$0.043 &   2 &	 9.2e-03 & 8 \\
\hline
   HD 127821 &    IRAS &   60 &   0.344$\pm$0.041 &   0.376$\pm$0.045 &   3 &	   0.011 & 8 \\
             &    IRAS &  100 &   0.525$\pm$0.137 &   0.529$\pm$0.138 &   2 &	 3.8e-03 & 8 \\
\hline
   HD 130693 &    IRAS &   60 &   0.149$\pm$0.042 &   0.154$\pm$0.043 & ... &	 3.2e-03 & 8 \\
\hline
   HD 131835 &    IRAS &   25 &   0.186$\pm$0.034 &   0.224$\pm$0.040 &   2 &	 6.1e-03 & 8 \\
             &    IRAS &   60 &   0.684$\pm$0.062 &   0.681$\pm$0.061 &   3 &	 1.0e-03 & 8 \\
\hline
   HD 157728 &    IRAS &   25 &   0.213$\pm$0.019 &   0.221$\pm$0.020 &   3 &	   0.053 & 8 \\
             &    IRAS &   60 &   0.536$\pm$0.048 &   0.531$\pm$0.048 &   3 &	 9.0e-03 & 8 \\
\hline
   HD 158352 &    IRAS &   25 &   0.258$\pm$0.044 &   0.255$\pm$0.043 &   3 &	   0.075 & 8 \\
             &    IRAS &   60 &   0.366$\pm$0.062 &   0.348$\pm$0.059 &   2 &	   0.013 & 8 \\
\hline
   HD 164249 &    IRAS &   60 &   0.647$\pm$0.091 &   0.669$\pm$0.094 &   2 &	 4.7e-03 & 8 \\
             & ISOPHOT &   60 &   0.740$\pm$0.037 &   0.761$\pm$0.038 & ... &	 4.7e-03 & 1 \\
             & ISOPHOT &   90 &   0.568$\pm$0.040 &   0.560$\pm$0.039 & ... &	 2.1e-03 & 1 \\
\hline
   HD 169666 &    MIPS &   24 &   0.087$\pm$0.009 &   0.090$\pm$0.009 & ... &	   0.039 & 7 \\
             &    IRAS &   25 &   0.101$\pm$0.013 &   0.088$\pm$0.011 &   3 &	   0.036 & 8 \\
\hline
   HD 170773 &    IRAS &   60 &   0.504$\pm$0.086 &   0.555$\pm$0.094 &   2 &	 9.1e-03 & 8 \\
             & ISOPHOT &   60 &   0.547$\pm$0.027 &   0.562$\pm$0.028 & ... &	 9.1e-03 & 1 \\
             & ISOPHOT &   90 &   0.771$\pm$0.054 &   0.821$\pm$0.057 & ... &	 4.0e-03 & 1 \\
\hline
   HD 172555 &    IRAS &   12 &    1.52$\pm$0.06  &    1.34$\pm$0.05  &   3 &	   0.508 & 8 \\
             &    IRAS &   25 &    1.09$\pm$0.06  &    0.91$\pm$0.05  &   3 &	   0.119 & 8 \\
             &    IRAS &   60 &   0.306$\pm$0.046 &   0.241$\pm$0.036 &   3 &	   0.020 & 8 \\
\hline
   HD 181296 &    IRAS &   12 &   0.481$\pm$0.024 &   0.351$\pm$0.018 &   3 &	   0.264 & 8 \\
             &    IRAS &   25 &   0.491$\pm$0.025 &   0.496$\pm$0.025 &   3 &	   0.061 & 8 \\
             &    IRAS &   60 &   0.495$\pm$0.045 &   0.449$\pm$0.040 &   3 &	   0.011 & 8 \\
             & ISOPHOT &   60 &   0.436$\pm$0.022 &   0.433$\pm$0.022 & ... &	   0.011 & 1 \\
             & ISOPHOT &   90 &   0.307$\pm$0.022 &   0.286$\pm$0.021 & ... &	  4.6e-03 &1 \\
\hline
   HD 181327 &    IRAS &   25 &   0.248$\pm$0.020 &   0.286$\pm$0.023 &   3 &	   0.027 & 8 \\
             &    IRAS &   60 &    1.86$\pm$0.07  &    1.93$\pm$0.08  &   3 &	  4.7e-03 & 8 \\
             & ISOPHOT &   60 &    1.69$\pm$0.17  &    1.73$\pm$0.17  & ... &	 4.7e-03 & 1 \\
             & ISOPHOT &   90 &    1.40$\pm$0.14  &    1.41$\pm$0.14  & ... &	 2.1e-03 & 1 \\
             &    IRAS &  100 &    1.72$\pm$0.21  &    1.69$\pm$0.20  &   2 &	 1.7e-03 & 8 \\
             & ISOPHOT &  170 &   0.820$\pm$0.214 &   0.736$\pm$0.192 & ... &	 5.9e-04 & 1 \\
\hline
   HD 182681 &    IRAS &   60 &   0.412$\pm$0.066 &   0.426$\pm$0.068 &   2 &	 5.6e-03 & 8 \\
\hline
   HD 191089 &    IRAS &   25 &   0.339$\pm$0.051 &   0.387$\pm$0.058 &   2 &	   0.024 & 8 \\
             &    IRAS &   60 &   0.711$\pm$0.057 &   0.729$\pm$0.058 &   3 &	 4.1e-03 & 8 \\
             & ISOPHOT &   60 &   0.774$\pm$0.080 &   0.781$\pm$0.081 & ... &	 4.1e-03 & 1 \\
             & ISOPHOT &   90 &   0.394$\pm$0.040 &   0.370$\pm$0.038 & ... &	 1.8e-03 & 1 \\
\hline
   HD 192758 &    MIPS &   24 &   0.042$\pm$0.004 &   0.041$\pm$0.004 & ... &      0.022 & 7 \\
             &    IRAS &   60 &   0.360$\pm$0.050 &   0.388$\pm$0.054 & ... &	 3.5e-03 & 8 \\
             & ISOPHOT &   60 &   0.370$\pm$0.022 &   0.383$\pm$0.023 & ... &	 3.5e-03 & 1 \\
             & ISOPHOT &   90 &   0.375$\pm$0.026 &   0.384$\pm$0.027 & ... &	 1.6e-03 & 1 \\
\hline
   HD 197481 &    IRAS &   60 &   0.269$\pm$0.046 &   0.252$\pm$0.043 &   2 &	   0.024 & 8 \\
             &    MIPS &   70 &   0.196$\pm$0.015 &   0.207$\pm$0.016 & ... &	   0.017 & 2 \\
             &   SHARC &  350 &   0.072$\pm$0.020 &   0.072$\pm$0.020 & ... &	 6.8e-04 & 2 \\
             &   SCUBA &  850 &   0.014$\pm$0.002 &   0.014$\pm$0.002 & ... &	 1.1e-04 & 6  \\
\hline
   HD 202917 & ISOPHOT &   60 &   0.046$\pm$0.014 &   0.047$\pm$0.014 & ... &	 1.9e-03 & 1 \\
\hline
   HD 205674 &    IRAS &   60 &   0.199$\pm$0.040 &   0.205$\pm$0.041 &   3 &	 3.7e-03 & 8 \\
\hline
   HD 206893 &    IRAS &   60 &   0.228$\pm$0.039 &   0.247$\pm$0.042 &   3 &	 6.3e-03 & 8 \\
             & ISOPHOT &   60 &   0.234$\pm$0.015 &   0.242$\pm$0.016 & ... &	 6.3e-03 & 1 \\
             & ISOPHOT &   90 &   0.259$\pm$0.018 &   0.268$\pm$0.019 & ... &	 2.8e-03 & 1 \\
\hline
   HD 218396 &    IRAS &   60 &   0.410$\pm$0.061 &   0.450$\pm$0.068 &   3 &	 8.6e-03 & 8 \\
             & ISOPHOT &   60 &   0.400$\pm$0.020 &   0.412$\pm$0.021 & ... &	 8.6e-03 & 1 \\
             & ISOPHOT &   90 &   0.553$\pm$0.039 &   0.585$\pm$0.041 & ... &	 3.8e-03 & 1 \\
\hline
   HD 221853 &    MIPS &   24 &   0.079$\pm$0.008 &   0.082$\pm$0.008 & ... &	   0.019 & 7 \\
             &    IRAS &   60 &   0.327$\pm$0.062 &   0.329$\pm$0.062 &   3 &	 3.0e-03 & 8 \\
             &     ISO &   60 &   0.368$\pm$0.031 &   0.376$\pm$0.032 & ... &	 3.0e-03 & 1 \\
             &     ISO &   90 &   0.231$\pm$0.016 &   0.223$\pm$0.015 & ... &	 1.3e-03 & 1 \\
\enddata
\tablecomments{Col.(1): Names. Col.(2): Instrument. 
 Col.(3): Wavelength. Cols.(4-5): Measured flux density and uncertainty at the specific wavelength. 
 On the infrared flux density values in Col.(5) color correction was applied. No color correction was applied on
 submillimeter fluxes.
 Col.(6): Quality of flux density, if available.
 Col.(7): Predicted flux density of the stellar photosphere at 
 the specific wavelength. For a detailed description of the prediction method, see Sect.~\ref{selection}.  
 Col.(8): References for papers related to the quoted flux and its uncertainty in Cols.(4-5).
}
\tablenotetext{a}{No quoted flux uncertainty in the IRAS Serendipitous Survey Catalog.
We assumed $\delta F_{25} = 0.15$\,Jy, and $\delta F_{60} = 0.32$\,Jy (4\% relative uncertainty in both cases).}
\tablerefs{(1) \'Abrah\'am et al., 2006, in prep.; (2) Chen et al. (2005b); (3) Greaves et al. (2000);
(4) {\sl http://data.spitzer.caltech.edu/popular/feps/20051123\_enhanced\_v1/}, 
 FEPS Data Explanatory Supplement v3.0, Hines et al. (2005); 
(5) Kleinmann et al. (1986); (6) Liu et al. (2004); (7) Mo\'or et al. (2006), in prep.; (8) Moshir et al. (1989); 
(9) Rieke et al. (2005); (10) Williams et al. (2004). 
}
\end{deluxetable}

%%%%%%%%%%%%%%%%%%%%%%%%%%%%%%%%%%%%%%%%%%%%%%%%%%%%%%%%%%%%%%%%%%%%%%%%%%%%%%%%%%%%%%%%%%%%%%%%%%%%%%%%%%%%%%%%%%%%%%
%%%%%%%%%%%%%%%%%%%%%%%%%%%%%%%%%%%%%%%%%%%%%%%%%%%%%%%%%%%%%%%%%%%%%%%%%%%%%%%%%%%%%%%%%%%%%%%%%%%%%%%%%%%%%%%%%%%%%%

\begin{deluxetable}{llccccccl} 
\tabletypesize{\scriptsize}
\tablecaption{Radial velocity information and membership in moving groups\label{tab3}}
\tablehead{
\colhead{Name} & \colhead{$\rm V_{rad}$} & Ref. & \colhead{U} & \colhead{V} & 
\colhead{W} & \colhead{Assoc. SKG} & \colhead{Prob.} & Ref. \\
%\colhead{} & \colhead{velocity} & \colhead{} & \colhead{} & \colhead{} & 
%\colhead{} & \colhead{SKG} & \colhead{probability} \\
\colhead{} & \colhead{[km~s$^{-1}$]} &  & \colhead{[km~s$^{-1}$]} & \colhead{[km~s$^{-1}$]} & 
\colhead{[km~s$^{-1}$]} & \colhead{} & \colhead{[\%]} &
}
\startdata
              HD 105  & ~$\rm +1.6\pm0.3$    & 7 &~$\rm -9.8 \pm 0.4 $ &$\rm -21.5 \pm  0.8 $ &~$\rm   -1.3\pm 0.3 $ & $\rm TucHor$     & 78& 6 \\
              HD 377  & ~$\rm +1.3\pm0.3$    & 7 &$\rm -14.4 \pm 0.6 $ &~$\rm  -7.0 \pm  0.4 $ &~$\rm   -3.9\pm 0.3 $ &... 	      &... & 15\\
             HD 3003  & ~$\rm +7.0\pm2.0$    & 13&~$\rm -9.6 \pm 0.8 $  &$\rm  -20.9 \pm  1.0$ &~$\rm  -0.6\pm 1.6 $  & $\rm TucHor$      & 85& 15\\ 
	     HD 6798  & $\rm +10.0\pm4.3$    & 5 &$\rm -36.6 \pm 2.7 $ &$\rm  -13.5 \pm  3.5$ &~$\rm  +4.1\pm 1.3 $  &...		&... &... \\
	     HD 8907  & ~$\rm +8.8\pm0.4$    & 7 &$\rm -10.4 \pm 0.3 $ &~$\rm  -4.6 \pm  0.4 $ &$\rm  -16.7\pm 0.4 $ &...		&... &... \\
             HD 9672  & $\rm +11.4\pm1.8$    & 5 &$\rm -23.9 \pm 1.1 $ &$\rm -16.6 \pm  0.8 $ &~$\rm   -6.6\pm 1.7 $ &...		&... &... \\
	    HD 10472  & $\rm +19.5\pm1.6$    & this work &~$\rm -8.5  \pm 0.7 $ &$\rm -24.2 \pm  1.1 $ &$\rm  -10.4\pm 1.3 $ & $\rm LA$	         & 38& this work \\
            HD 10647  & $\rm +27.5\pm0.2$    & 7 &~$\rm -1.2  \pm 0.1 $ &$\rm -26.6 \pm  0.2 $ &$\rm  -17.8\pm 0.2 $ &...		 &... &... \\
            HD 10638  & ~$\rm -0.4\pm1.2$    & 11&$\rm -12.3 \pm 1.0 $ &$\rm -25.0 \pm  1.5 $ &$\rm  -14.8\pm 1.0 $ & $\rm LA $         & 29& this work \\
            HD 15115  & ~$\rm +8.8\pm3.0$    & this work &$\rm -13.2 \pm 1.9 $ &$\rm -17.8 \pm  1.2 $ &~$\rm   -6.0\pm 2.3 $ & $\rm BPMG$        & 22& this work \\
            HD 15745  & $\rm +10.5\pm1.2$    & this work &$\rm -16.5 \pm 1.1 $ &$\rm -10.8 \pm  1.3 $ &$\rm  -10.7\pm 0.7 $ &...		 &... &... \\
            HD 16743  & $\rm +21.9\pm1.1$    & this work &$\rm -23.6 \pm 0.9 $ &$\rm -18.0 \pm  0.6 $ &$\rm  -15.1\pm 0.9 $ &...		 &... &... \\
	    HD 17390  & ~$\rm +7.2\pm1.8$    & 7 &$\rm -15.3 \pm 0.8 $ &~$\rm  -9.6 \pm  0.5 $ &~$\rm   +1.0\pm 1.6 $ &...		 &... &... \\
            HD 21997  & $\rm +17.3\pm0.8$    & 5 &$\rm -12.9 \pm 0.5 $ &$\rm -22.3 \pm  1.0 $ &~$\rm   -3.9\pm 0.9 $ & $\rm GAYA2$	 & 62& this work \\
            HD 24966  & $\rm +29.6\pm2.2$    & this work &$\rm -15.0 \pm 0.9 $ &$\rm -26.4 \pm  1.4 $ &$\rm  -13.2\pm 1.8 $ &...		 &... &... \\
            HD 25457  & $\rm +17.6\pm0.2$    & 7 &~$\rm -7.9 \pm 0.2 $  &$\rm -28.7 \pm  0.4 $ &$\rm  -11.9\pm 0.2 $ & $\rm ABDor$       & 30& 16 \\
	    HD 30447  & $\rm +21.3\pm2.5$    & this work &$\rm -13.1 \pm 1.4 $ &$\rm -20.9 \pm  1.6 $ &~$\rm   -3.8\pm 1.7 $ & $\rm GAYA2$       & 54& this work \\
            HD 32297  & $+21.8\pm2.1$        & this work &$\rm -16.3\pm2.0  $  &$\rm -16.0 \pm 1.5$   & $\rm  -11.0\pm 1.0 $ & ...		 &... &... \\ 
	    HD 35841  & $\rm +23.1\pm1.3$    & this work &$\rm -13.0 \pm 1.0 $ &$\rm -21.3 \pm  1.7 $ &~$\rm   -3.6\pm 1.7 $ & $\rm GAYA2$       & 56& this work \\
            HD 37484  & $\rm +23.0\pm2.6$    & this work &$\rm -11.6 \pm 1.4 $ &$\rm -20.4 \pm  1.9 $ &~$\rm   -5.2\pm 1.3 $ & $\rm TucHor$      & 25& this work \\
            HD 38207  & $\rm +24.9\pm1.4$    & this work &$\rm -14.7 \pm 1.1 $ &$\rm -21.2 \pm  1.5 $ &~$\rm   -4.1\pm 1.5 $ & $\rm GAYA2$       & 32& this work \\
            HD 38206  & $\rm +24.9\pm0.6$    & 5 &$\rm -13.7 \pm 0.5 $ &$\rm -21.2 \pm  0.5 $ &~$\rm   -6.1\pm 0.4 $ & $\rm GAYA2$       & 27& this work \\
	    HD 38678  & $\rm +18.9\pm2.7$    & 5 &$\rm -13.6 \pm 2.0 $ &$\rm -10.5 \pm  1.6 $ &~$\rm   -8.1\pm 1.0 $ & $\rm Castor$      & 28& 1 \\
	    HD 39060  & $\rm +20.2\pm0.4$    & 14&$\rm -10.9 \pm 0.1 $ &$\rm -16.2 \pm  0.3 $ &~$\rm   -9.2\pm 0.2 $ & $\rm BPMG$	 & 98& 2 \\
            HD 50571  & $\rm +22.2\pm1.2$    & 7 &$\rm -16.2  \pm 0.4 $&$\rm -22.3 \pm  1.1 $ &~$\rm  -4.4\pm 0.5  $ &...	         &... &... \\
	    HD 53143  & $\rm +21.9\pm0.1$    & 7 &$\rm -25.5 \pm 0.3 $ &$\rm -18.1 \pm  0.1 $ &$\rm  -15.2\pm 0.1 $ &...	         &... &... \\
            HD 54341  & $\rm +47.4\pm1.4$    & this work &$\rm -16.5 \pm 0.5 $ &$\rm -44.0 \pm  1.3 $ &~$\rm   -8.7\pm 0.5 $ &...	         &... &... \\
            HD 69830  & $\rm +29.8\pm0.1$    & 7 &$\rm +28.9 \pm 0.5 $ &$\rm -60.9 \pm  0.4 $ &$\rm  -10.1\pm 0.2 $ &...	         &... &... \\
	    HD 76582  & ~$\rm -4.0\pm2.5$    & 11&$\rm +10.7 \pm 1.8 $ &~$\rm +5.1 \pm  1.1  $ &~$\rm   +9.9\pm 1.5 $ &...		 &... &... \\
	    HD 78702  & $\rm +16.9\pm2.1$    & 5 &$\rm -27.5 \pm 1.6 $ &$\rm -10.5 \pm  1.8 $ &~$\rm   -7.3\pm 1.1 $ &...	         &... &... \\
            HD 84870A & ~$\rm +3.7\pm0.4$    & this work &~$\rm -7.8 \pm 0.5 $ &$\rm -25.5 \pm  2.0 $ &~$\rm   -5.7\pm 0.7 $ &...	         &... &... \\  
	    HD 85672  & ~$\rm -5.2\pm6.8$    & 4 &~$\rm -9.5  \pm 4.2 $ &~$\rm  -3.0 \pm  1.6 $ &$\rm  -14.8\pm 5.4 $ &...	         &... &... \\
            HD 92945  & $\rm +22.6\pm0.2$    & 7 &$\rm -15.2 \pm 0.3 $ &$\rm -27.8 \pm  0.2 $ &~$\rm   -4.3\pm 0.3 $ &...	         &... &... \\
           HD 107146  & ~$\rm +1.5\pm0.2$    & 7 &$\rm -10.6 \pm 0.3 $ &$\rm -28.8 \pm  0.7 $ &~$\rm   -5.2\pm 0.3 $ &...	       &... &... \\
           HD 109573A & ~$\rm +9.4\pm2.3$    & 5 &~$\rm -9.0  \pm 1.3 $ &$\rm -19.0 \pm  1.9 $ &~$\rm   -4.4\pm 1.0 $ & $\rm TWA$ 	& 64& 10\\
           HD 110058  & $\rm +21.7\pm1.3$    & this work &$\rm +0.0  \pm 1.3 $ &$\rm -26.6 \pm  1.4 $ &~$\rm   -2.1\pm 0.8 $ &...	         &... &... \\
	   HD 115116  & ~$\rm -2.3\pm0.7$    & 4 &$\rm -30.5 \pm 1.5 $ &$\rm -17.2 \pm  1.1 $ &~$\rm   -1.8\pm 0.3 $ &...		 &... &... \\
           HD 120534  & $\rm +46.7\pm1.1$    & this work &$\rm +21.8 \pm 1.8 $ &$\rm -43.1 \pm  3.2 $ &$\rm  +15.0\pm 1.8 $ &...		 &... &... \\
           HD 121812  & $\rm -15.2\pm0.7$    & 8 &$\rm -29.6 \pm 1.2 $ &$\rm -61.3 \pm  2.7 $ &~$\rm   -1.8\pm 0.9 $ &...		 &... &... \\
           HD 122106  & ~$\rm -1.6\pm5.0$    & 7 &~$\rm +0.6 \pm 2.6 $  &$\rm -21.7 \pm  2.0 $ &~$\rm   -9.0\pm 4.2 $ &...		 &... &... \\
	   HD 127821  & $\rm -15.4\pm2.6$    & 7 &$\rm -16.0 \pm 0.5 $ &$\rm -26.2 \pm  1.6 $ &~$\rm   -2.7\pm 2.0 $ &...		 &... &... \\
	   HD 130693  & $\rm +14.3\pm0.8$    & this work &~$\rm +8.1  \pm 0.9 $ &~$\rm  -7.3 \pm  0.6 $ &$\rm  +10.4\pm 0.8 $ &...		 &... &... \\
           HD 131835  & ~$\rm +3.3\pm1.7$    & this work &~$\rm -4.6  \pm 1.6 $ &$\rm -17.2 \pm  2.1 $ &~$\rm   -3.0\pm 1.0 $ & $\rm UCL$	 & 50& 12\\
           HD 157728  & $\rm -19.7\pm1.2$    & 11&~$\rm -7.0  \pm 0.8 $ &$\rm  -21.3 \pm  0.8$ &~$\rm  -4.4\pm 0.6 $  &...		 &... &... \\
	   HD 158352  & $\rm -36.1\pm1.2$    & 11&$\rm -35.5  \pm 1.1$ &$\rm  -19.9 \pm  0.5$ &~$\rm +7.4\pm 0.9 $   &...		 &... &... \\
	   HD 164249  & ~$\rm -0.2\pm0.5$    & 7 &~$\rm -7.6  \pm 0.6 $ &$\rm -15.3 \pm  0.7 $ &~$\rm   -8.9\pm 0.4 $ & $\rm BPMG$        & 20& 9 \\
           HD 169666  & $\rm -44.3\pm0.6$    & 7 &~$\rm +1.3  \pm 0.3 $ &$\rm  -46.8 \pm  0.6$ &~$\rm  -7.9\pm 0.5 $  &...	         &... &... \\
	   HD 170773  & $\rm -26.3\pm1.1$    & 7 &$\rm -30.5 \pm 1.1 $ &~$\rm  -3.8 \pm  0.2 $ &$\rm  -12.3\pm 0.6 $ &...	         &... &... \\
           HD 172555  & ~$\rm +2.0\pm2.5 $   & 11&$\rm -11.0 \pm 2.0 $ &$\rm -15.6 \pm  1.2 $ &~$\rm   -9.3\pm 1.0 $ & $\rm BPMG$        & 96& 9 \\
           HD 181296  & ~$\rm -2.0\pm10.0 $  & 13&$\rm -10.7 \pm 8.6 $ &$\rm -14.9 \pm  2.7 $ &~$\rm   -7.3\pm 4.4 $ & $\rm BPMG$        & 87& 9 \\
           HD 181327  & ~$\rm +0.2\pm0.4$    & 7 &~$\rm -9.1  \pm 0.5 $ &$\rm -16.2 \pm  0.7 $ &~$\rm   -8.4\pm 0.4 $ & $\rm BPMG$	& 57& 9 \\
           HD 182681  & ~$\rm +1.4\pm5.0$    & 3 &~$\rm -0.4  \pm 4.6 $ &$\rm  -13.4 \pm  1.1$ &$\rm  -10.8\pm 1.8 $ &...	       &... &... \\
	   HD 191089  & ~$\rm -5.8\pm2.2$    & 7 &~$\rm -7.8  \pm 1.9 $ &$\rm -16.2 \pm  0.9 $ &$\rm  -10.3\pm 1.2 $ & $\rm BPMG$	& 38& this work \\
	   HD 192758  & $\rm -11.1\pm1.2$    & this work &$\rm -18.4 \pm 2.1 $ &$\rm -13.8 \pm  2.8 $ &~$\rm   -6.7\pm 2.7 $ & $\rm IC2391$	 & 50& this work \\
           HD 197481  & ~$\rm -4.5\pm1.3$    & 5 &$\rm -10.1 \pm 1.0 $ &$\rm -16.4 \pm  0.3 $ &$\rm  -10.5\pm 0.8 $ & $\rm BPMG$       & 56& 2 \\
           HD 202917  & ~$\rm -1.6\pm0.2$    & 7 &~$\rm -8.2  \pm 0.4 $ &$\rm -20.0 \pm  1.1 $ &~$\rm   -0.3\pm 0.2 $ & $\rm TucHor$	& 48& 15\\
           HD 205674  & ~$\rm +1.1\pm5.1$    & this work &~$\rm -1.6  \pm 3.0 $ &$\rm -24.5 \pm  2.4 $ &$\rm  -16.6\pm 3.7 $ &...	       &... &... \\
           HD 206893  & $\rm -12.9\pm1.4$    & 7 &$\rm -20.0 \pm 0.9 $ &~$\rm  -7.7 \pm  0.7 $ &~$\rm   -2.1\pm 1.0 $ &...		 &... &... \\
           HD 218396  & $\rm -12.6\pm1.3$    & 5 &$\rm -12.4 \pm 0.5 $ &$\rm -21.4 \pm  1.1 $ &~$\rm   -7.4\pm 0.9 $ & $\rm LA  $	 & 62& this work \\
           HD 221853  & ~$\rm -4.2\pm2.1$    & this work &$\rm -12.7 \pm 0.9 $ &$\rm -20.6 \pm  1.8 $ &$\rm  -11.2\pm 1.9 $ & $\rm LA  $	 & 94& this work \\

\enddata
\tablecomments{Col.(1): Names. Col.(2): Radial velocity and its uncertainty. Col.(3): Reference for the source of measurement. 
Cols.(4-6): U,V,W Galactic space velocity components,U is positive towards 
the Galactic centre, V is positive in the direction of galactic rotation and W is positive towards the North galactic pole. 
Col.(7): Assigned stellar kinematic group (see Table\,4). Col.(8): Membership probability. 
Col.(9): Reference for membership identification. }
\tablerefs{(1) Barrado y Navascu\'es et al. (1998); (2) Barrado y Navascu\'es et al. (1999); (3) Evans (1967); 
(4) Grenier et al. (1999); 
(5) Kharchenko et al. and references therein (2004);
(6) Mamajek et al. (2004a); (7) Nordstr\"om et al. and references therein (2004); (8) Strassmeier et al. (2000); 
(9) Song et al. (2003); 
(10) Webb et al.(1999); (11) Wilson (1953); (12) de Zeeuw et al. (1999); (13) Zuckerman \& Webb (2000); 
(14) Zuckerman et al. (2001a);
(15) Zuckerman et al. (2001b); (16) Zuckerman \& Song (2004b)}
\end{deluxetable}

%%%%%%%%%%%%%%%%%%%%%%%%%%%%%%%%%%%%%%%%%%%%%%%%%%%%%%%%%%%%%%%%%%%%%%%%%%%%%%%%%%%%%%%%%%%%%%%%%%%%%%%%%%%%%%%%%%%%%%
%%%%%%%%%%%%%%%%%%%%%%%%%%%%%%%%%%%%%%%%%%%%%%%%%%%%%%%%%%%%%%%%%%%%%%%%%%%%%%%%%%%%%%%%%%%%%%%%%%%%%%%%%%%%%%%%%%%%%%

\begin{deluxetable}{lccccc} 
\tabletypesize{\scriptsize}
%\rotate
\tablecaption{Description of stellar kinematic groups \label{tab2}} 
\tablewidth{0pt}
\tablehead{
\colhead{Group Name} & \colhead{U,V,W} & Ref. & \colhead{Age} & Ref. \\
\colhead{}& \colhead{[km~s$^{-1}$]} & &   \colhead{[Myr]} & }
\startdata
AB Dor Moving Group                 & -8, -27, -14      & 13 & 50  & 13\\
                                    &                 &    & 100 &  3\\
$\beta$\,Pictoris Moving Group      & -11, -16, -9      & 13 & $\rm 12^{+8}_{-4}$ & 11\\
Castor Moving Group                 & -10.7, -8.0, -9.7 &  1 & $\rm 200\pm100 $  & 1 \\
Great Austral Young Association2    & -11.0, -22.5, -4.6&  9 & $\rm 20$ & 9\\
Hyades cluster                      & -40, -17, -3      & 13 & 600 & 13 \\
IC 2391 supercluster                & -20.6, -15.7, -9.1&  5 & 35--55 & 5 \\
Local Association                   & -11.6, -21.0, -11.4& 5 & 20--150& 5 \\
Lower Centaurus Crux                & -8.2, -18.6, -6.4  & 6 & $\rm 16\pm1$ & 4\\
Upper Centaurus Lupus               & -6.8, -19.3, -5.7  & 6 & $\rm 17\pm1$ & 4\\
Tucana/Horologium                   & -11, -21, 0        &13 & $\rm 30$& 13 \\
                                    &                  &   & 10--30& 7\\
                                    &                  &   & 20& 8 \\
                                    &                  &   & 10--40 & 12 \\
TW Hydrae Association               & -11, -18, -5       & 13& 8 & 13\\
                                    &                  &   &5--15 & 10\\
Ursa Major Moving Group             & +14, +1, -9        & 13& 300 & 13\\
                                    &                  &   & 500 & 2\\
\enddata
\tablerefs{(1) Barrado y Navascu\'es et al. (1998); (2) King et al. (2003); (3) Luhman et al. (2005); 
(4) Mamajek et al. (2002); (5) Montes et al. and references therein (2001); (6) Sartori et al. (2003);
(7) Stelzer \& Neuh\"auser (2000); (8) Torres et al. (2000); (9) Torres et al. (2002); 
(10) Weintraub et al. (2000); (11) Zuckerman et al.(2001a); (12) Zuckerman et al. (2001b); 
(13) Zuckerman \& Song and references therein (2004b);
}

\end{deluxetable}

%%%%%%%%%%%%%%%%%%%%%%%%%%%%%%%%%%%%%%%%%%%%%%%%%%%%%%%%%%%%%%%%%%%%%%%%%%%%%%%%%%%%%%%%%%%%%%%%%%%%%%%%%%%%%%%%%%%%%%
%%%%%%%%%%%%%%%%%%%%%%%%%%%%%%%%%%%%%%%%%%%%%%%%%%%%%%%%%%%%%%%%%%%%%%%%%%%%%%%%%%%%%%%%%%%%%%%%%%%%%%%%%%%%%%%%%%%%%%

\begin{deluxetable}{lcc} 
\tabletypesize{\scriptsize}
%\rotate
\tablecaption{Comparison of age estimates for newly identified moving group
members \label{tabage}} 
\tablewidth{0pt}
\tablehead{
\colhead{Name} & \colhead{Age estimates} & \colhead{References} \\
\colhead{} & \colhead{[Myr]} & \colhead{}
}
\startdata
HD 10472    & $\rm 2000^{+1000}_{-1400}$  &  5 \\
            & 30                          &  8 \\
	    & [20,150]                    &  this study \\
\tableline	    
HD 15115    & $\rm 900^{+1300}_{-900}$    &  5 \\
            & 100                         &  8 \\
	    & $\rm 12^{+8}_{-4}$          &  this study \\
\tableline
HD 21997    & 100                         &  8 \\
            & $\rm 20^{+10}_{-10}$        &  this study \\
\tableline	    
HD 30447    & $\rm 2100^{+700}_{-700}$    &  5 \\
            & $\lesssim 100$              &  8 \\
	    & $\rm 20^{+10}_{-10}$        &  this study \\
\tableline	    
HD 37484    & $\rm 1500^{+900}_{-1500}$   &  5 \\
            & [30,100]                    &  1 \\ 
	    & $\rm 30^{+10}_{-20}$        &  this study \\
\tableline	    
HD 38207    & [100,300]                   &  1 \\
            &  $\rm 20^{+10}_{-10}$       &  this study \\
\tableline	    
HD 38206    & $\sim \rm 9^{+14}_{-9}$     &  3 \\
            &  $\rm 20^{+10}_{-10}$       &  this study \\
\tableline	    
HD 191089   & $\rm 17^{+8}_{-4}$          &  4 \\
            & $\rm 3000^{+700}_{-900}$    &  5 \\
	    & $\lesssim 100$              &  8 \\
	    & $\rm 12^{+8}_{-4}$          &  this study \\
\tableline	    
HD 218396   & $\rm 732^{+396}_{-682}$     &  7 \\
            & 30                          &  8 \\
	    & [20,150]                    &  this study \\
\tableline	    
HD 221853   & 800                         &  6 \\
            & 1800                        &  2 \\
	    & $\rm1700^{+500}_{-600}$     &  5 \\
	    &$\lesssim 100$               &  8 \\
	    & [20,150]                    &  this study \\
\enddata
\tablerefs{(1) Carpenter et al. (2005); (2) Decin et al. (2000);
(3) Gerbaldi et al. (1999); (4) Mamajek (2004); (5) Nordstr\"om et al. (2004);
(6) Silverstone (2000); (7) Song et al. (2001); (8) Zuckerman \& Song (2004a);
}
\end{deluxetable}

%%%%%%%%%%%%%%%%%%%%%%%%%%%%%%%%%%%%%%%%%%%%%%%%%%%%%%%%%%%%%%%%%%%%%%%%%%%%%%%%%%%%%%%%%%%%%%%%%%%%%%%%%%%%%%%%%%%%%%
%%%%%%%%%%%%%%%%%%%%%%%%%%%%%%%%%%%%%%%%%%%%%%%%%%%%%%%%%%%%%%%%%%%%%%%%%%%%%%%%%%%%%%%%%%%%%%%%%%%%%%%%%%%%%%%%%%%%%%

\begin{deluxetable}{llccccc} 
\tabletypesize{\scriptsize}
%\rotate
\tablecaption{List of bogus disks \label{tabbogus}} 
\tablewidth{0pt}
\tablehead{
\colhead{Name} & \colhead{Nearby IRAS} & \colhead{First reference} & \colhead{Reason} & \colhead{Instrument} & \multicolumn{2}{c}{Position of unrelated} \\
\colhead{} & \colhead{Source} & \colhead{as debris disk} & \colhead{of rejection} & \colhead{} & \multicolumn{2}{c}{nearby IR source} \\
\colhead{} & \colhead{} & \colhead{candidate} & \colhead{} & \colhead{} &\colhead{RA(2000)} & \colhead{DEC(2000)}

}
\startdata
HD 23484  &  F03423--3826  & 2\tablenotemark{a} & no detectable excess\tablenotemark{a}  & ISOPHOT &  ... & ...\\
HD 34739  &  F05154--5301 & 2 & source confusion  & MIPS & 5 16 36.3 & -52 57 40\\
HD 53842  &  F06539--8355 & 1 & source confusion\tablenotemark{b} & MIPS &  6 46 02.8 & -83 59 37 \\
HD 56099  &  F07149+5913  & 2 & source confusion  & MIPS  & 7 19 09.6 & +59 07 20 \\   
HD 72390  &  F08210--8414 & 2 & source confusion  & ISOPHOT  & 08 14 39 & -84 23 23 \\
HD 82821  &  F09319+0346  & 2 &	source confusion  & MIPS & 09 34 36.2 & +03 32 39 \\
HD 143840 &  F16001--0440 & 2 & extended emission & ISOPHOT, MIPS & ... & ... \\
HD 158373 &  F17265--0957  & 2\tablenotemark{a} & no detectable excess\tablenotemark{a}  & ISOPHOT &  ... & ...\\
HD 164330 &  F17559+6236  & 2\tablenotemark{a} & no detectable excess\tablenotemark{a}  & ISOPHOT &  ... & ...\\
HD 185053 &  F19415--8123 & 2 & extended emission & MIPS  & ... & ...\\
HD 204942 &  F21297--2422 & 2 & source confusion  & MIPS  & 21 32 35.8  & -24 09 30\\
\enddata
\tablecomments{Col.(1): Name. Col.(2): Identification of nearby IRAS source. The IRAS source 
is always located within 30$''$ of the star position. 
 Col.(3): Reference for first mention as debris disk.
 Col.(4): Reason why the object was classified as bogus disk and rejected from further analyses. 
 Col.(5): Instrument. Cols.(6-7): 
 When the object was nominated as bogus disk due to source confusion, position of the unrelated nearby IR source is given. 
 In the cases of MIPS observations MOPEX was used to extract source coordinates from 70$\mu m$ MIPS images. 
 The coordinates of peak brightness in ISOPHOT maps were determined by fitting a point source profile.}
\tablenotetext{a}{Silverstone (2000) selected as debris disk candidate on the basis of IRAS data, but found
that ISOPHOT observations did not confirm the detection of IR excess. He proposed that the non-detection
of far-infrared flux excess towards the star's position can be explained by cirrus contamination.}
\tablenotetext{b}{HD\,53842 is not really bogus. At 24$\mu$m it shows IR excess above the photosphere 
(see Mo\'or et al. 2006, in prep.).
However at 70$\mu$m the excess emission is related to a nearby infrared source. 
} 
\tablerefs{(1) Mannings \& Barlow (1998);
(2) Silverstone (2000).
}
\end{deluxetable}

%%%%%%%%%%%%%%%%%%%%%%%%%%%%%%%%%%%%%%%%%%%%%%%%%%%%%%%%%%%%%%%%%%%%%%%%%%%%%%%%%%%%%%%%%%%%%%%%%%%%%%%%%%%%%%%%%%%%%%
%%%%%%%%%%%%%%%%%%%%%%%%%%%%%%%%%%%%%%%%%%%%%%%%%%%%%%%%%%%%%%%%%%%%%%%%%%%%%%%%%%%%%%%%%%%%%%%%%%%%%%%%%%%%%%%%%%%%%%

\begin{deluxetable}{llccc} 
\tabletypesize{\scriptsize}
%\rotate
\tablecaption{List of rejected suspicious objects \label{tabsusp}} 
\tablewidth{0pt}
\tablehead{
\colhead{Name} & \colhead{Nearby IRAS} & \colhead{First reference} & \colhead{Reason of} & \colhead{Name of} \\
\colhead{} & \colhead{Source} & \colhead{as debris disk} & \colhead{suspicion} & \colhead{background source} \\
\colhead{} & \colhead{} & \colhead{candidate} & \colhead{} & \colhead{}
}
\startdata
HIP 13005 &  F02444+1505 & 2 & nearby extended 2MASS source\tablenotemark{a} & 2MASS J02471368+1518315 (XSC 122165)   \\ 
HD 33095  &  F05049--1927 & 2 & cirrus                       & ...                         \\
HD 36162  &  F05275+1519 & 2 & nearby extended 2MASS source   & 2MASS J05303093+1521513 (XSC 2620748)  \\
HD 39944  &  F05526-2535 & 1 & nearby galaxy\tablenotemark{b} & ESO 488-41                  \\
HD 83870  &  F09393+4111 & 2 & nearby galaxy 	             & PGC 27759                   \\
HD 97455  &  F11107+5541 & 2 & nearby galaxy	             & PGC 34197                   \\
HD 124718 &  F14129--2707 & 2 & nearby 2MASS source with an    & 2MASS J14155109-2721050            \\
         &              &   & excess in the $\rm K_s$ band   &                             \\
HD 140775 &  F15429+0536 & 2 & star locates in the wall of    & ...                         \\ 
         &              &   & Local Bubble or beyond         &                             \\
HD 154145 &  F17011--0004 & 2 & star locates in the wall of    & ...                         \\ 
         &              &  & Local Bubble or beyond	     &                             \\
\enddata 
\tablecomments{Col.(1): Name. Col.(2): Identification of nearby IRAS source. The IRAS source 
is always located within 30$''$ of the star position. 
 Col.(3): Reference for first mention as debris disks.
 Col.(4): Reason of suspicion. For more details see Sect.~\ref{rejection}. 
 Col.(5): Name of background source.}
\tablenotetext{a}{\citet{zs04} also found suspicious this debris candidate based on the offset
between its IRAS positions measured at 12 and 60$\mu m$.}
\tablenotetext{b}{\citet{sylvester00} also noted this coincidence between the position of the IRAS source and 
the nearby galaxy.}
\tablerefs{(1) Mannings \& Barlow (1998);
(2) Silverstone (2000).
}
\end{deluxetable}

%%%%%%%%%%%%%%%%%%%%%%%%%%%%%%%%%%%%%%%%%%%%%%%%%%%%%%%%%%%%%%%%%%%%%%%%%%%%%%%%%%%%%%%%%%%%%%%%%%%%%%%%%%%%%%%%%%%%%%
%%%%%%%%%%%%%%%%%%%%%%%%%%%%%%%%%%%%%%%%%%%%%%%%%%%%%%%%%%%%%%%%%%%%%%%%%%%%%%%%%%%%%%%%%%%%%%%%%%%%%%%%%%%%%%%%%%%%%%

\clearpage

\begin{figure} 
\epsscale{1.}
\plotone{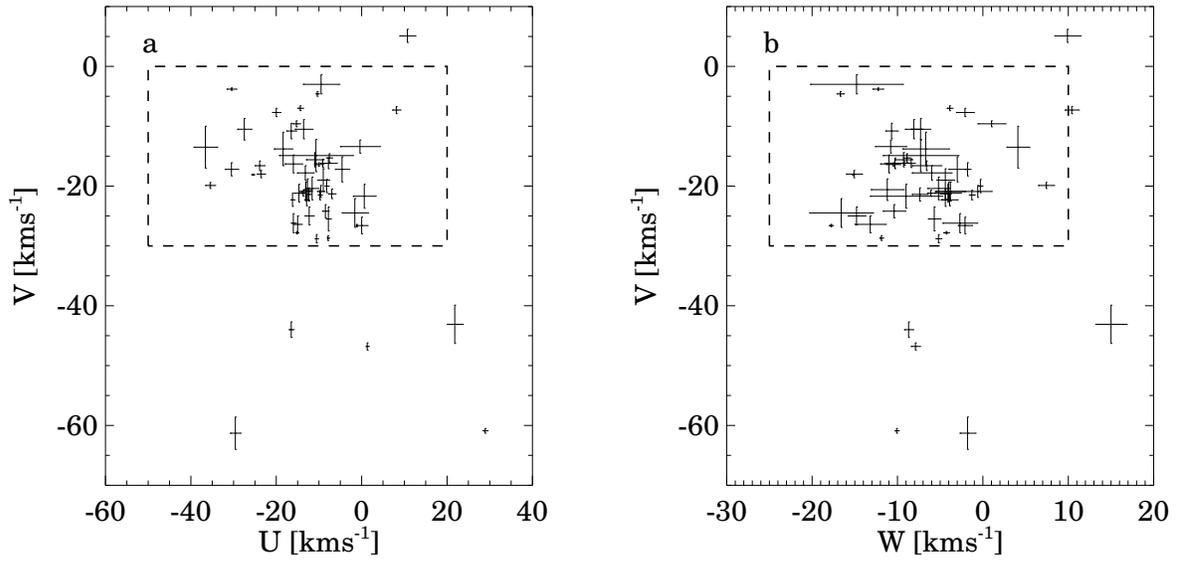}
\caption{U,V and W,V planes for stars in Table~\ref{tab3}. The dashed rectangle marks the young disk 
population as defined by \citet{leggett92}. \label{figuvw}}
\end{figure}

%\begin{figure} 
%\epsscale{.8}
%\plotone{f5.eps}
%\caption{Same as Fig.~\ref{fdage1} but for disks whose existence had been 
%proved independently of IRAS.
%\label{fdage2}}
%\end{figure}

\begin{thebibliography}{}
\bibitem[\'Abrah\'am et al.(2003)]{ap03} \'Abrah\'am, P., Mo\'or, A., Kiss, Cs., H\'eraudeau, P., \& del Burgo, C., 
Exploiting the ISO Data Archive. Infrared Astronomy in the Internet Age, held in Siguenza, Spain 24-27 June, 2002. 
Edited by C. Gry, S. Peschke, J. Matagne, P. Garcia-Lario, R. Lorente, \& A. Salama. Published as ESA Publications Series, 
ESA SP-511. European Space Agency, 2003, p. 129.
\bibitem[Ardila et al.(2005)]{ardila05} Ardila, D. R., Golimowski, D. A., Krist, J. E., et al. 2004, ApJ, 617, L147
\bibitem[Artymowicz(1996)]{artym96} Artymowicz, P. 1996, in The Role of Dust in the Formation of Stars, ed.
H. U. Ka\"ufl \& R. Siebenmorgen (New York: Springer), 137
\bibitem[Aumann et al.(1984)]{aumann84} Aumann, H. H., Beichman, C. A., Gillett, F. C. et al., 1984, \apj, 278, L23
\bibitem[Backman \& Gillett(1987)]{backman87} Backman, D., \& Gillett, F. C. 1987, in Cool Stars, Stellar Systems and
the Sun, ed. J. L. Linsky \& R. E. Stencel (Lecture Notes in Physics 291; Berlin: Springer), 340
\bibitem[Backman \& Paresce(1993)]{backman93} Backman, D. E., \& Paresce, F. 1993, in Protostars and Planets III, ed.
E. H. Levy \& J. I. Lunine (Tucson: Univ. Arizona Press), 1253
\bibitem[Barrado y Navascu\'es(1998)]{barrado98} Barrado y Navascu\'es, D. 1998, \aap, 339, 831
\bibitem[Barrado y Navascu\'es et al.(1999)]{barrado99} Barrado y Navascu\'es, D., Stauffer, J. R., Song, I., \& Caillault, J.-P. 1999,
\apj, 520, L123 
\bibitem[Beichman et al.(1988)]{beichman98} IRAS Explanatory Supplement, 1988, ed. Beichman, C. A., Neugebauer, G., Habing, H. J., Clegg, P. E.,
\& Chester T. J. (Washington, DC: US GPO)
\bibitem[Beichman et al.(2005a)]{beichman05a} Beichman, C. A., Bryden, G., Rieke, G. H., et al. 2005a, ApJ, 622, 1160
\bibitem[Beichman et al.(2005b)]{beichman05b} Beichman, C. A., Bryden, G., Gautier, T. N., et al. 2005b, \apj, 626, 1061
\bibitem[Binney et al.(1997)]{binney97} Binney J. J., Dehnen W., Houk N., Murray C. A., Penston M. J., 1997, 
Proc. ESA Symp., Hipparcos - Venice '97. ESA SP-402, 473
\bibitem[Bryden et al.(2005)]{bryden05} Bryden, G., Beichman, C. A., Trilling, D. E., et al. 2005, astro-ph/0508165
\bibitem[Carpenter et al.(2005)]{carpenter05} Carpenter, J.M., Wolf, S., Schreyer, K., Launhardt R., \& Henning T. 
2005, \aj, 129, 1049 
\bibitem[Chen et al.(2005a)]{chen05a} Chen, C. H., Jura, M., Gordon, K. D., \& Blaylock, M. 2005a, \apj, 623, 493
\bibitem[Chen et al.(2005b)]{chen05b} Chen, C. H., Patten, B. M., Werner, W., et al. 2005b, ApJ preprint doi:10.1086/'497124'
\bibitem[Cutri et al.(2003)]{cutri03} Cutri, R. M., et al., 2003, Explanatory Supplement to the 2MASS All Sky Data Release (Pasadena: IPAC)
\bibitem[Decin et al.(2000)]{decin00} Decin, G., Dominik, C., Malfait, K., Mayor, M., \& Waelkens, C. 2000, \aap, 357, 533
\bibitem[Decin et al.(2003)]{decin03} Decin, G., Dominik, C., Waters, L. B. F. M., \& Waelkens, C. 2003,
\apj, 598, 636
\bibitem[de la Reza \& Pinz\'on(2004)]{delareza04} de la Reza, R. \& Pinz\'on, G. 2004, \aj, 128, 1812
\bibitem[Dent et al.(2000)]{dent00} Dent, W. R. F., Walker, H. J., Holland, W. S., \& Greaves, J. S. 2000, \mnras, 314, 702
\bibitem[Dominik \& Decin(2003)]{dominik03} Dominik, C., \& Decin, G. 2003, \apj, 598, 626
\bibitem[Eggen(1989)]{eggen89} Eggen, O. J. 1989, PASP, 101, 54
\bibitem[ESA(1997)]{esa97} ESA, 1997, The Hipparcos and Tycho Catalogues, ESA SP-1200
\bibitem[Evans(1967)]{evans67} Evans, D. S. 1967,
 Determination of Radial Velocities and their Applications, Proceedings from IAU Symposium no. 30 held at the 
 University of Toronto 20-24 June, 1966. Edited by Alan Henry Batten and John Frederick Heard. 
 International Astronomical Union. Symposium no. 30, Academic Press, London, p.57
\bibitem[Favata et al.(1993)]{favata93} Favata, F., Barbera, M., Micela, G., \& Sciortino, S. 1993, \aap, 277, 428
\bibitem[Gabriel et al.(1997)]{gabriel97} Gabriel, C., Acosta-Pulido, J., Heinrichsen, I., Morris, H., \& Tai, W.-M. 1997, 
The ISOPHOT Interactive Analysis PIA, a calibration and scientific analysis tool, in Astronomical Data Analysis Software 
and Systems (ADASS) VI, ed. G. Hunt, \& H. E. Payne (San Francisco: ASP), ASP Conf. Ser., 125, 108
\bibitem[Gerbaldi et al.(1999)]{gerbaldi99} Gerbaldi, M., Faraggiana, R., Burnage, R., Delmas, F., G\'omez, A. E., \&
Grenier, S. 1999, A\&AS, 137, 273
\bibitem[Girardi et al.(2000)]{girardi00} Girardi, L., Bressan, A., Bertelli, G., \& Chiosi, C. 2000, A\&AS, 141, 371
\bibitem[Gordon et al.(2005)]{gordon05} Gordon, K. D., Rieke, G., Engelbracht C., et al. 2005, \pasp, 117, 503
\bibitem[Greaves et al.(2000)]{greaves00} Greaves, J. S., Mannings, V., \& Holland, W. S. 2000, Icarus, 143, 155 
\bibitem[Greenberg et al.(1998)]{greenberg98} Greenberg, J. M. 1998, \aap, 330, 375
\bibitem[Grenier et al.(1999)]{grenier99} Grenier, S., Baylac, M. O., Rolland, L., et al. 1999, A\&AS, 137, 451
\bibitem[Habing et al.(2001)]{habing01} Habing, H. J., Dominik, C., Jourdain de Muizon, M. et al. 2001, \aap, 365, 545
\bibitem[Harper et al.(1984)]{harper84} Harper, D. A., Loewenstein, R. F., \& Davidson, J. A. 1984, \apj, 285, 808
\bibitem[Heinrichsen et al.(1999)]{heinrichsen99} Heinrichsen, I., Walker, H. J., Klaas, U., Sylvester, R. J., 
 \& Lemke, D. 1999, MNRAS, 304, 589
\bibitem[Helou \& Walker(1988)]{helou} IRAS Catalogs: The Small Scale Structure Catalog, 1988, ed. G. Helou and D.W. Walker, NASA RP-1190, vol 7 (Washington, DC: GPO)
\bibitem[Hines et al.(2005)]{hines05} Hines, D.C. et al. 2005, "FEPS Data Explanatory Supplement," Version 3.0, (Pasadena:
SSC).
\bibitem[Holland et al.(1998)]{holland98} Holland, W., Greaves, J., Zuckerman, B., et al. 1998, Nat, 392,
788
\bibitem[Jarrett et al.(2000)]{jarrett00} Jarrett, T. H., Chester, T., Cutri, R., et al. 2000, \aj, 120, 298
%\bibitem[Jayawardhana et al.(2001)]{jay01} Jayawardhana, R., Fisher, R. S., Telesco, C. M. et al. 2001, \aj, 122, 2047
\bibitem[Jura(1991)]{jura91} Jura, M. 1991, \apj, 383, L79
\bibitem[Jura et al.(1998)]{jura98} Jura, M., Malkan, M., White, R., et al. 1998, \apj, 505, 897 
\bibitem[Jura et al.(2004)]{jura04} Jura, M., Chen, C. H., Furlan, E., et al. 2004, ApJS, 154, 453
\bibitem[Kalas et al.(2002)]{kalas02} Kalas, P., Graham, J. R., Beckwith, S. V. W., Jewitt, D. C., \& Lloyd, J. P. 2002, \apj, 567, 999
\bibitem[Kalas et al.(2005)]{kalas05a} Kalas, P., Graham, J. R., \& Clampin, M. 2005, Nature, 435, 1067
\bibitem[Kalas(2005)]{kalas05b} Kalas, P. 2005, astro-ph/0511244
\bibitem[Kenyon(2002)]{kenyon02a} Kenyon, S. J. 2002, PASP, 114, 265
\bibitem[Kenyon \& Bromley(2002)]{kenyon02} Kenyon, S. J., \& Bromley, B. C. 2002, \apj, 577, L35
\bibitem[Kenyon \& Bromley(2004)]{kenyon04} Kenyon, S. J., \& Bromley, B. C. 2004, \apj, 127, 513
\bibitem[Kessler et al.(1996)]{kessler96} Kessler, M. F., Steinz, J. A., Anderegg, M. F., et al. 1996, \aap, 315, L27
\bibitem[Kharchenko(2004)]{kharchenko04} Kharchenko N.V., Piskunov A.E., Scholz R.-D. 2004, Astron. Nachr., 325, 439
\bibitem[Kim et al.(2005)]{kim05} Kim, J. S, Hines, D. C., Backman, D. E., et al. 2005, \apj, 632, 659
\bibitem[King et al.(2003)]{king03} King, J. R., Villarreal, A. R., Soderblom, D. R., 
Gulliver, A. F., \& Adelman, S. J. 2003, \aj, 125, 1980
\bibitem[Kleinmann et al.(1986)]{kleinmann86} Kleinmann, S.G., Cutri, R.M., Young, E.T., Low, F.J., \& Gillett, F.C. 1986, 
Explanatory Supplement to the IRAS Serendipitous Survey Catalog (Pasadena: JPL).
\bibitem[Koerner et al.(1998)]{koerner98} Koerner, D. W., Ressler, M. E., Werner, M. W., \& Backman, D. E. 1998, \apj, 503, L83
\bibitem[Krist et al.(2005)]{krist05} Krist, J. E., et al. 2005, \aj, 129, 1008
\bibitem[Lachaume et al.(1999)]{lachaume99} Lachaume, R., Dominik, C., Lanz, T., \& Habing, H. J. 1999,\aap, 348, 897 
\bibitem[Lallement et al.(2003)]{lall03} Lallement, R., Welsh, B. Y., Vergely, J. L., Crifo, F., \& Sfeir, D. 2003, \aap, 411, 447
\bibitem[Laureijs et al.(2003)]{laureijs03} Laureijs, R. J., Klaas, U., Richards, P. J., Schulz, B., \& \'Abrah\'am, P. 2003, 
The ISO Handbook Vol. IV.: PHT - The Imaging Photo-Polarimeter, Version 2.0.1, ESA SP-1262, European Space Agency
\bibitem[Leggett(1992)]{leggett92} Leggett, S. K. 1992, ApJS, 82, 351
\bibitem[Lemke et al.(1996)]{lemke96} Lemke, D., Klaas, U., Abolins, J., et al. 1996, \aap, 315, L64
\bibitem[Lisse et al.(2002)]{lisse} Lisse, C., et al. 2002, ApJ, 570, 779
\bibitem[Liu et al.(2004)]{liu04} Liu, M. C., Matthews, B. C., Williams, J. P., \& Kalas, P. G. 2004, \apj, 608, 526 
\bibitem[Low et al.(2005)]{low05} Low, F. J., Smith, P. S., Werner, M., et al. 2005, \apj, 631, 1170
\bibitem[Lowrance et al.(2000)]{lowrance00} Lowrance, P. J., Schneider, G., Kirkpatrick, J. D., et al. 2000, \apj, 541, 390
\bibitem[Luhman et al.(2005)]{luhman05} Luhman, K. L., Stauffer, John R., Mamajek, E. E. 2005, \apj, 628, L69
\bibitem[Magakian(2003)]{magakian03} Magakian, T. Y. 2003, \aap, 399, 141 
\bibitem[Makovoz \& Marleau(2005)]{makovoz05} Makovoz, D., \& Marleau, F. 2005, \pasp, 117, 1113 
\bibitem[Mamajek et al.(2002)]{mamajek02} Mamajek, E. E., Meyer, M. R., Liebert, J. 2002, \aj, 124, 1617 
\bibitem[Mamajek et al.(2004a)]{mamajek04} Mamajek, E. E., Meyer, M. R., Hinz, P. M., et al. 2004a, \apj, 612, 496
\bibitem[Mamajek (2004b)]{mamajek04b} Mamajek, E. E., 2004b, PhD Thesis, The University of Arizona 
\bibitem[Mannings \& Barlow(1998)]{mb98} Mannings, V., \& Barlow, M. 1998, \apj, 497, 330
\bibitem[Meyer et al.(2004)]{meyer04} Meyer, M. R., Hillenbrand, L. A., Backman, D. E., et al. 2004, ApJS, 154, 422
\bibitem[Meynet et al.(1993)]{meynet93} Meynet, G., Mermilliod, J.-C., \& Maeder, A. 1993, A\&AS, 98, 477
\bibitem[Montes et al.(2001)]{montes01} Montes, D., et al. 2001, \mnras, 328, 45 
\bibitem[Moshir et al.(1989)]{moshir89} Moshir, M., et al. 1989, Explanatory Supplement to the IRAS Faint Source Survey (Pasadena: JPL) (FSC)
%\bibitem[Najita \& Williams(2005)]{najita05} Najita, J., Williams J. P. 2005, astro-ph/0508165
\bibitem[Nordstr\"om et al.(2004)]{nord04} Nordstr\"om, B., et al. 2004, \aap, 418, 989
\bibitem[Oudmaijer et al.(1992)]{oudmaijer92} Oudmaijer, R. D., van der Veen, W. E. C. J.,
 Waters, L. B. F. M., et al. 1992, A\&AS, 96, 625
\bibitem[Patten \& Willson(1991)]{pw91} Patten, B. M., \& Willson, L. A. 1991, \aj, 102, 323
\bibitem[Plets \& Vynckier(1999)]{plets99} Plets, H., \& Vynckier, C. 1999, \aap, 343, 496
\bibitem[Rebull et al.(2004)]{rebull04} Rebull, L. M., Stapelfeldt, K. R., Chen, C., et al.
2004, AAS, 205, 1703
\bibitem[Rieke et al.(2004)]{rieke04} Rieke, G. H., et al. 2004, ApJS, 154, 25
\bibitem[Rieke et al.(2005)]{rieke05} Rieke, G. H., et al. 2005, \apj, 620, 1010
\bibitem[Rocha-Pinto et al.(2004)]{rp04} Rocha-Pinto, H. J., Flynn, C., Scalo, J. et al. 2004, \aap, 423, 517
\bibitem[Sadakane \& Nishida(1986)]{sn86} Sadakane, K., \& Nishida, M. 1986, \pasp, 98, 685
\bibitem[Sartori et al.(2003)]{sartori03} Sartori, M. J., L\'epine, \& J. R. D., Dias, W. D. 2003, 
\aap, 404, 913 
\bibitem[Schneider et al.(1999)]{schneider99} Schneider, G., et al. 1999, ApJ, 513, L127 
\bibitem[Schneider et al.(2005)]{schneider05} Schneider, G., Silverstone, M. D., \& Hines, D. C. 2005, \apj, 629, 117L
\bibitem[Silverstone(2000)]{sil00} Silverstone, M. D. 2000, Ph.D. thesis, UCLA
\bibitem[Skuljan et al.(1999)]{skuljan99} Skuljan, J., Hearnshaw, J. B., \& Cottrell, P. L. 1999, \mnras, 308, 731
\bibitem[Smith \& Terrile(1984)]{smith84} Smith, B., \& Terrile, R. 1984, Sci, 226, 1421
\bibitem[Song et al.(2000)]{song00} Song, I., Caillault, J.-P., Barrado y Navascu\'es, D., Stauffer, J. R., \& Randich, S. 2000, 
\apj, 533, L41
\bibitem[Song et al.(2001)]{song01} Song, I., Caillault, J.-P., Barrado y Navascu\'es, D., \&  Stauffer, J. R. 2001, 
\apj, 546, 352
\bibitem[Song et al.(2003)]{song03} Song, I., Zuckerman, B., \& Bessell, M. S. 2003, \apj, 599, 342
\bibitem[Song et al.(2004)]{song04} Song, I., Zuckerman, B., \& Bessell, M. S. 2004, \apj, 614, L125
\bibitem[Spangler et al.(2001)]{sprangler01} Spangler, C., Sargent, A. I., Silverstone, M. D., Becklin, E. E., \&
Zuckerman, B. 2001, \apj, 555, 932
\bibitem[Stauffer et al.(1998)]{stauffer98} Stauffer, J.R., Schultz, G., \& Kirkpatrick, J.D., 1998, \apj, 499, L199
\bibitem[Stauffer et al.(2005)]{stauffer05} Stauffer, J.R., Rebull, L. M., Carpenter, J. et al. 2005, \aj, 130, 1834
\bibitem[Stelzer \& Neuhäuser(2000)]{stelzer2000} Stelzer, B., \& Neuh\"auser, R. 2000, A\&A, 361, 581
\bibitem[Stencel \& Backman(1991)]{sb91} Stencel, R. E.,\& Backman, D. E. 1991, \apjs, 75, 905
\bibitem[Strassmeier et al.(2000)]{strassmeier00} Strassmeier, K., Washuettl, A., Granzer, Th., Scheck, M., \& Weber, M.
 2000, A\&AS, 142, 275
\bibitem[Sylvester \& Mannings(2000)]{sylvester00} Sylvester, R., Mannings, V. 2000, MNRAS, 313, 73
\bibitem[Thi et al.(2001)]{thi01} Thi, W. F., et al. 2001, ApJ, 561, 1074
\bibitem[Torres et al.(2000)]{torres00} Torres, C. A. O., da Silva, L., Quast, G. R., de la Reza, R., \& Jilinski, E. 2000, AJ, 120, 1410 
\bibitem[Torres et al.(2002)]{torres02} Torres, C. A. O., Quast, G. R., de la Reza, R., da Silva, L., 
\& Melo, C. H. F. 2002, astro-ph/0207078
\bibitem[Uzpen et al.(2005)]{uzpen05} Uzpen, B., Kobulnicky, H. A., Olsen, K. A. G., et al. 2005, \apj, 629, 512
\bibitem[Walker \& Heinrichsen(2000)]{walker00} Walker, H.J., \& Heinrichsen, I. 2000, Icarus, 143, 147
\bibitem[Webb et al.(1999)]{webb99} Webb, R. A., Zuckerman, B., Platais, I., et al. 1999, \apj, 512, L63
\bibitem[Weintraub et al.(2000)]{weintraub00} Weintraub, D. A., Saumon, D., Kastner, J. H., \& Forveille, T. 2000, \apj, 530, 867
\bibitem[Werner et al.(2004)]{werner04} Werner, M. W., et al. 2004, ApJS, 154, 1
\bibitem[Williams et al.(2004)]{williams04} Williams, J. P., Najita, J., Liu, M. C., et al. 2004, \apj, 604, 414 
\bibitem[Wilson(1953)]{wilson53} Wilson R.E., 1953, General Catalogue of Stellar Radial Velocities, 
Carnegie Inst. of Washington Publ. 601, Washington DC
\bibitem[Wright et al.(2003)]{wright03} Wright C.O., Egan M.P., Kraemer K.E., Price S.D., 2003, \aj, 125, 359 
\bibitem[Wright et al.(2004)]{wright04} Wright, J. T., Marcy, G. W., Butler, R. P., \& Vogt, S. S. 2004, \apj, 152, 261
\bibitem[Wyatt(2005)]{wyatt05} Wyatt, M. C. 2005, \aap, 433, 1007
\bibitem[de Zeeuw et al.(1999)]{deZeeuw99} de Zeeuw, P. T., Hoogerwerf, R., de Bruijne, J. H. J., Brown, A. G. A., Blaauw, A. 1999,
\aj, 117, 354 
\bibitem[Zuckerman \& Webb(2000)]{zw00} Zuckerman, B., \& Webb, R. A. 2000, \apj, 535, 959
\bibitem[Zuckerman(2001)]{zuckerman01} Zuckerman, B. 2001, ARA\&A, 39, 549
\bibitem[Zuckerman et al.(2001a)]{zs01} Zuckerman, B., Song, I., Bessel, M. S. \& Webb, R. A. 2001a, \apj, 562, L87
\bibitem[Zuckerman et al.(2001b)]{zs01c} Zuckerman, B., Song, I., Webb, R. A. 2001b, \apj, 559, 388
\bibitem[Zuckerman \& Song(2004a)]{zs04} Zuckerman, B., \& Song, I. 2004a, \apj, 603, 738
\bibitem[Zuckerman \& Song(2004b)]{zs04b} Zuckerman, B., \& Song, I. 2004b, ARA\&A, 42, 685

\end{thebibliography}
\end{document}